\documentclass[10pt, journal, compsoc]{IEEEtran}

\usepackage[pdftex, svgnames, dvipsnames]{xcolor}
\usepackage[colorlinks]{hyperref}
\usepackage{xspace}
\usepackage{amsmath}
\usepackage{cleveref}
\usepackage{amssymb}
\usepackage{mathabx}
\usepackage{centernot}
\usepackage{booktabs}
\usepackage{fixme}
\usepackage[acronym, nomain]{glossaries}
\usepackage{tikz}
\usepackage[lined]{algorithm2e}
\usepackage{etoolbox}
\usepackage{pgfplots}
\pgfplotsset{compat=1.17}

\usepackage{tabularx}
\usepackage{multirow}
\usepackage{fontawesome}
\usepackage{tablefootnote}
\usepackage{makecell}

\newcommand{\attackstep}{k}
\newcommand{\concreteattackstep}{\overline{k}}
\newcommand{\instance}{p}
\newcommand{\part}{a}
\newcommand{\requirement}{r}
\newcommand{\state}{T}

\newcommand{\dpinst}[2]{#1(#2)}
\newcommand{\attackpath}{K}
\newcommand{\concreteattackpath}{\overline{K}}
\newcommand{\attackpathlist}{\overline{\attackpath}}
\newcommand{\po}[2]{\left[ #1, #2 \right]}
\newcommand{\protection}{P}
\newcommand{\solution}{S}

\newcommand{\dpinsts}{\mathbb{D}}
\newcommand{\asset}{\alpha} 
\newcommand{\assets}{\mathbb{A}}

\newcommand{\artifacts}{\mathcal{A}}
\newcommand{\concreteattackpaths}{\overline{\attackpaths}}
\newcommand{\attackpaths}{\attacksteps} 
\newcommand{\attacksteps}{\mathcal{K}}
\newcommand{\pos}{\mathcal{O}}
\newcommand{\protections}{\mathcal{P}}
\newcommand{\requirements}{\mathcal{R}}
\newcommand{\states}{\mathcal{T}}
\newcommand{\solutions}{\mathcal{S}}

\DeclareMathOperator{\joint}{\sqcap}
\DeclareMathOperator{\notpreceded}{\nprec}
\DeclareMathOperator{\preceded}{\prec}
\DeclareMathOperator{\shouldnotpreceded}{\preceded^-}
\DeclareMathOperator{\shouldpreceded}{\preceded^+}

\DeclareMathOperator{\compatible}{compatible}
\DeclareMathOperator{\cyclomatic}{cyclomatic}
\DeclareMathOperator{\enforce}{protect}
\DeclareMathOperator{\guardedins}{\instructions.guarded}
\DeclareMathOperator{\halstead}{halstead}
\DeclareMathOperator{\heaviside}{H}
\DeclareMathOperator{\instructions}{instructions}
\DeclareMathOperator{\likelihood}{\Lambda}
\DeclareMathOperator{\measure}{measure}
\DeclareMathOperator{\metric}{predict}
\DeclareMathOperator{\overhead}{overhead}
\DeclareMathOperator{\pindex}{index}
\DeclareMathOperator{\potency}{\mathfrak{P}}

\DeclareMathOperator{\localins}{\instructions.local}
\DeclareMathOperator{\remoteins}{\instructions.remote}
\DeclareMathOperator{\weight}{weight}
\DeclareMathOperator{\art}{art}

\newcommand{\maxapis}{\sigma}
\newcommand{\maxoverhead}{\theta}
\newcommand{\measuremalus}{\rho}
\newcommand{\measureweight}{\tau}
\newcommand{\minmeasure}{\epsilon}
\newcommand{\mitigation}{\zeta}
\newcommand{\probability}{\pi}
\newcommand{\synergy}{\omega}

\newcommand{\ccs}[2]{#1^{\left\{#2\right\}}}
\newcommand{\false}{\bot}
\newcommand{\nonnegativereals}{\reals_{\ge 0}}
\newcommand{\reals}{\mathbb{R}}
\newcommand{\true}{\top}
\newcommand{\vanilla}{\varnothing}

\fxsetup{status=draft, layout=inline, theme=color}
\FXRegisterAuthor{bds}{brds}{\colorbox{Purple}{\color{White}BDS}}
\FXRegisterAuthor{ab}{aab}{\colorbox{Brown}{\color{White}AB}}
\FXRegisterAuthor{bc}{ebc}{\colorbox{Red}{\color{White}BC}}
\FXRegisterAuthor{lr}{alr}{\colorbox{Orange}{\color{White}LR}}
\FXRegisterAuthor{dc}{adc}{\colorbox{Periwinkle}{\color{White}DC}}

\newcommand{\ie}{i.e.\@\xspace}
\newcommand{\eg}{e.g.\@\xspace}
\newcommand{\etal}{\emph{~et~al.\@}\xspace}

\glsdisablehyper
\newacronym{po}{PO}{Protection Objective}
\newcommand{\PO}{\gls{po}\xspace}
\newacronym{dsp}{DSP}{Deployed Software Protection}
\newacronym{csp}{CSP}{Concrete Software Protection}
\newacronym{cs}{CS}{Candidate Solution}
\newacronym{asp}{ASP}{Abstract Software Protection}
\newcommand{\DSP}{\gls{dsp}\xspace}
\newcommand{\DSPs}{\glspl{dsp}\xspace}
\newcommand{\ASP}{\gls{asp}\xspace}
\newcommand{\ASPs}{\glspl{asp}\xspace}
\newcommand{\CSP}{\gls{csp}\xspace}
\newcommand{\CSPs}{\glspl{csp}\xspace}

\newcommand{\CSs}{\glspl{cs}\xspace}
\newacronym{poset}{poset}{partially ordered set}

\newacronym{ccs}{CCS}{Code Correlation Set}
\newcommand{\CCS}{\gls{ccs}\xspace}
\newcommand{\CCSs}{\glspl{ccs}\xspace}

\LinesNumbered
\DontPrintSemicolon
\SetAlCapSty{}

\SetCommentSty{commentfont}

\SetNlSty{numberfont}{}{}
\AtBeginEnvironment{algorithm}{\SetKwSty{textsc}}
\SetKw{Break}{break}
\SetKw{Continue}{continue}
\newcommand{\algorithmname}[2][]{\textsc{#2}\textsubscript{\textsc{#1}}\xspace}

\newcommand{\Explore}[1][]{\algorithmname[#1]{Explore}}
\newcommand{\nil}{\algorithmname{nil}}

\newacronym{as}{AS}{Attack Step}

\newacronym{ap}{AP}{Attack Path}
\newacronym{va}{VA}{Vanilla Application}
\newacronym{pa}{PA}{Protected Application}
\newacronym{aa}{AA}{Application Artifact}
\newacronym[plural=PIs,firstplural=Protection Indices (PIs)]{pi}{PI}{Protection Index}

\newcommand{\VA}{\gls{va}\xspace}

\renewcommand{\AA}{\gls{aa}\xspace}
\newcommand{\AAs}{\glspl{aa}\xspace}

\newcommand{\POs}{\glspl{po}\xspace}

\newacronym{nist}{NIST}{National Institute of Standards and Technology}\newcommand{\nist}{\gls{nist}\xspace}
\newacronym{gdpr}{GDPR}{General Data Protection Regulation}

\newacronym{aspire}{ASPIRE}{Advanced Software Protection: Integration, Research and Exploitation}
\newacronym{accl}{ACCL}{Client-side Communication Logic}
\newacronym{ascl}{ASCL}{Server-side Communication Logic}
\newacronym{actc}{ACTC}{ASPIRE Compiler Tool Chain}\newcommand{\actc}{\gls{actc}\xspace}
\newacronym{esp}{ESP}{Expert system for Software Protection}\newcommand{\esp}{\gls{esp}\xspace}
\newacronym{it}{IT}{Information Technology}\newcommand{\itech}{\gls{it}\xspace}

\newacronym{diablo}{DIABLO}{Diablo Is A Better Link-time Optimizer}
\newacronym{cff}{CFF}{Control Flow Flattening}
\newacronym{ip}{IP}{Intellectual Property}
\newacronym{drm}{DRM}{Digital Rights Management}\newcommand{\drm}{\gls{drm}\xspace}
\newacronym{cve}{CVE}{Common Vulnerabilities and Exposures}
\newacronym{cwe}{CWE}{Common Weakness Enumeration}
\newacronym{mate}{MATE}{Man-At-The-End}\newcommand{\mate}{\gls{mate}\xspace}
\newacronym{sp}{SP}{Software Protection}
\newcommand{\softprot}{\gls{sp}\xspace}
\newcommand{\SP}{\gls{sp}\xspace}
\newcommand{\SPs}{\glspl{sp}\xspace}
\newacronym{aslr}{ASLR}{Address Space Layout Randomization}
\newacronym{txt}{TXT}{Trusted eXecution Technology}
\newacronym{tpm}{TPM}{Trusted Platform Module}
\newacronym{tcg}{TCG}{Trusted Computing Group}
\newacronym{sgx}{SGX}{Software Guard Extensions}

\newacronym{dec}{DEC}{Digital Equipment Corporation}
\newacronym{vax}{VAX}{Virtual Address eXtension}

\newacronym{dendral}{DENDRAL}{DENDritic ALgorithm}
\newacronym{xcon}{XCON}{eXpert CONfigurer}
\newacronym{ides}{IDES}{Intrusion Detection Expert System}
\newacronym{nides}{NIDES}{Next-generation Intrusion Detection Expert System}
\newacronym{nidx}{NIDX}{Network Intrusion Detection eXpert system}
\newacronym{nadir}{NADIR}{Network Anomaly Detection and Intrusion Reporter}
\newacronym{audes}{AudES}{Expert System for security Auditing}

\newacronym{cfg}{CFG}{Control Flow Graph}
\newacronym{os}{OS}{Operating System}
\newacronym{dll}{DLL}{Dynamic-Link Library}
\newacronym{gnu}{GNU}{GNU is Not Unix}
\newacronym{gcc}{GCC}{\gls{gnu} C Compiler}
\newacronym{gdb}{GDB}{\gls{gnu} DeBugger}
\newacronym{foss}{FOSS}{Free and Open Source Software}
\newacronym{cdt}{CDT}{C/C++ Development Tooling}
\newacronym{ast}{AST}{Abstract Syntax Tree}
\newacronym{gui}{GUI}{Graphical User Interface}
\newacronym{xml}{XML}{eXtensible Mark-up Language}

\newacronym{sloc}{SLOC}{Source Lines Of Code}
\newacronym{hl}{HL}{Halstead's Length}
\newacronym{cc}{CC}{Cyclomatic Complexity}

\newacronym{cpu}{CPU}{Central Processing Unit}
\newacronym{fpu}{FPU}{Floating Point Unit}
\newacronym{alu}{ALU}{Arithmetic and Logic Unit}
\newacronym{gpu}{GPU}{Graphics Processing Unit}
\newacronym{gpgpu}{GPGPU}{General-Purpose computing on Graphics Processing Units}

\newacronym{lisp}{LISP}{LISt Processor}
\newacronym{cil}{CIL}{C Intermediate Language}

\newacronym{smb}{SMB}{Server Message Block}

\newacronym{mitm}{MITM}{Man-In-The-Middle}
\newacronym{ddos}{DDoS}{Distributed Denial of Service}

\newacronym{rat}{RAT}{Remote Access Trojan}

\newacronym{ids}{IDS}{Intrusion Detection System}
\newacronym{cysemol}{CySeMoL}{Cyber Security Modeling Language}
\newacronym{camel}{CAMEL}{Cloud Application Modelling \& Execution Language}
\newacronym{ml}{ML}{Machine Learning}
\newacronym{ai}{AI}{Artificial Intelligence}
\newacronym{som}{SOM}{Self-Organizing Maps}
\newacronym{ann}{ANN}{Artificial Neural Network}

\newacronym{milp}{MILP}{Mixed Integer-Linear Programming}

\newacronym{owl}{OWL}{Web Ontology Language}
\newacronym{owltwo}{OWL2}{\acrlong{owl} 2}

\newacronym{roi}{ROI}{Return In Investment}
\newacronym{byod}{BYOD}{Bring Your Own Device}
\newacronym{bsa}{BSA}{BSA | The Software Alliance}
\newacronym{otp}{OTP}{One-Time Password}
\newacronym{pin}{PIN}{Personal Identification Number}
\newacronym{sdlc}{SDLC}{Software Development Life Cycle}
\newacronym{kb}{KB}{Knowledge Base}\newcommand{\KB}{\gls{kb}\xspace}
\newacronym{poc}{PoC}{Proof of Concept}\newcommand{\poc}{\gls{poc}\xspace}
\newacronym{sto}{StO}{Security-through-Obscurity}

\usetikzlibrary{arrows.meta,shapes.multipart,positioning,calc}

\tikzstyle{class} =
[
	rectangle,
	draw = Dandelion!50!Black,
	semithick,
	text centered,
	anchor = north,
	text = black,
	shading = axis,
	bottom color = Dandelion!30,
	top color = Dandelion!5,
	shading angle = 45,
	rounded corners = 0.5mm,
	minimum width = 24mm,
	minimum height = 6mm,
	inner sep = 1mm,
	font = \sffamily\fontsize{8}{8}\selectfont
]
\tikzstyle{generic} =
[
	double,
	bottom color = Red!30,
	top color = Red!5
]
\tikzstyle{usecase} =
[
	double,
	bottom color = RoyalBlue!30,
	top color = RoyalBlue!5
]
\tikzstyle{connector} =
[
	draw = Black!85,
	text = Black!85,
	fill = Black!85,
	rounded corners,
	semithick
]
\newcommand{\UMLSimpleClass}[2][]
{
	\node[class, #1] (#2) {\textbf{#2}};
}
\newcommand{\UMLClass}[3][]
{
	\node[class, rectangle split, rectangle split parts = 2, rectangle split part align = {center, left}, #1] (#2)
	{
		\textbf{#2}\nodepart{second}\shortstack[l]{#3}
	};
}
\newcommand{\UMLSimpleEnumeration}[2][]
{
	\node[class, minimum height = 9mm, #1] (#2)
	{
		\shortstack{$\ll$enumeration$\gg$\\\textbf{#2}}
	};
}
\newcommand{\UMLEnumeration}[3][]
{
	\node[class, rectangle split, rectangle split parts = 2, rectangle split part align = {center, left}, minimum height = 9mm, #1] (#2)
	{
		\shortstack{$\ll$enumeration$\gg$\\\textbf{#2}}\nodepart{second}\shortstack[l]{#3}
	};
}
\newcommand{\UMLAssociation}[4][]
{
	\draw[connector, -{Straight Barb[angle' = 45, scale = 2]}, #1] (#2) -- node[inner sep = 0.5mm, fill = White, rounded corners = 0] {\sffamily\fontsize{8}{8}\selectfont#4} (#3);
}

\newcommand{\UMLAssociationRCorner}[4][]
{
	\draw[connector, fill = none, -{Straight Barb[angle' = 45, scale = 2]}, #1] (#2) -| node[inner sep = 0.5mm, fill = White, rounded corners = 0] {\sffamily\fontsize{8}{8}\selectfont#4} (#3);
}
\newcommand{\UMLComposition}[3][]{\draw[connector, -{Diamond[scale = 1.5]}, #1] (#2) -- (#3);}
\newcommand{\UMLGeneralization}[3][]{\draw[connector, -{Triangle[open, scale = 1.5]}, #1] (#2) -- (#3);}

\newcommand{\UMLAssociationLoop}[3][0]
{
	\draw[connector, fill = none, -{Straight Barb[angle' = 45, scale = 2]}] ($(#2.east) + (0mm, #1 * -1mm)$) -- ($(#2.east) + (-#1 * 2mm + 4mm, #1 * -1mm)$) |- node[inner sep = 0.5mm, fill = White, rounded corners = 0] {\sffamily\fontsize{8}{8}\selectfont#3} ($(#2.south) + (#1 * 2mm, #1 * 2mm - 4mm)$) -- ($(#2.south) + (#1 * 2mm, 0mm)$);
}

\definecolor{mygreen}{rgb}{0,0.6,0}
\definecolor{mygray}{rgb}{0.5,0.5,0.5}
\definecolor{mymauve}{rgb}{0.58,0,0.82}
\definecolor{codegray}{rgb}{0.5,0.5,0.5}

\begin{document}
\title{
Automatic Selection of Protections to Mitigate Risks Against Software Applications
}
\author{Daniele Canavese,
        Leonardo Regano,
        Bjorn De~Sutter,~\IEEEmembership{Member,~IEEE,}
        and~Cataldo Basile,~\IEEEmembership{Member,~IEEE}%
\IEEEcompsocitemizethanks{
\IEEEcompsocthanksitem D.~Canavese is with Institut de Recherche en Informatique de Toulouse. E-mail: daniele.canavese@irit.fr
\IEEEcompsocthanksitem L.~Regano is with the Dipartimento di Ingegneria Elettrica ed Elettronica, Universit\`a degli Studi di Cagliari, Cagliari, Italy. E-mail: leonardo.regano@unica.it
\IEEEcompsocthanksitem B.~De~Sutter is with the Computing Systems Lab of Ghent University. E-mail: bjorn.desutter@ugent.be
\IEEEcompsocthanksitem C.~Basile is with the Dipartimento di Automatica e Informatica, Politecnico di Torino, Torino, Italy. E-mail: cataldo.basile@polito.it
}
\thanks{Corresponding Author: L.~Regano}
\thanks{This work was partially supported by project SERICS (PE00000014) under the NRRP MUR program funded by the EU - NGEU, by ICO, Institut Cybersécurité Occitanie, funded by Région Occitanie, France, by the European research project Horizon Europe DUCA (GA 101086308), by the European FP7 research project ASPIRE (GA 609734) and CNRS IRN EU-CHECK.}
}

\maketitle

\fbox{\parbox{0.9\linewidth}{This work has been submitted to the IEEE for possible publication. Copyright may be transferred without notice, after which this version may no longer be accessible.}}\medskip

\begin{abstract}

This paper introduces a novel approach for the automated selection of software protections to mitigate \acrlong{mate} risks against critical assets within software applications. We formalize the key elements involved in protection decision-making---including code artifacts, assets, security requirements, attacks, and software protections--- and frame the protection process through a game-theoretic model. 
In this model, a defender strategically applies protections to various code artifacts of a target application, anticipating repeated attack attempts by adversaries {against the confidentiality and integrity of the application's assets}. The selection of the optimal defense maximizes resistance to attacks while ensuring the application remains usable by constraining the overhead introduced by protections. The game is solved through a heuristic based on a mini-max depth-first exploration strategy, augmented with dynamic programming optimizations for improved efficiency. 
Central to our formulation is the introduction of the Software Protection Index, an original contribution that extends existing notions of potency and resilience by evaluating protection effectiveness against attack paths using software metrics and expert assessments. 
We validate our approach through a proof-of-concept implementation and expert evaluations, demonstrating that automated software protection is a practical and effective solution for risk mitigation in software.

\end{abstract}

\begin{IEEEkeywords}
software protection, Man-at-the-End attacks, software risk mitigation, software potency and resilience
\end{IEEEkeywords}

\IEEEpeerreviewmaketitle

\section{Introduction}
Software impacts many aspects of our lives these days.
The business of companies developing software or creating or managing content and services with software  depends to a large degree on the resistance of the software against so-called \mate attacks~\cite{collberg-falcarin}.

{In the \mate attack model, attackers have full access to the software and complete control over the systems on which they aim to reverse engineer software and tamper with it to breach the security requirements of its assets. They can use various tools like simulators, debuggers, disassemblers, and decompilers. 
\mate attacks include reverse engineering (e.g., to steal algorithms or find vulnerabilities), tampering (e.g., to bypass license checks or cheat in games), or unauthorized execution (e.g., to run multiple copies with a single license).}

Defenders can only rely on protections within the software or remote trusted servers to mitigate the \mate risks against their software. 
Hence, {\softprot} refers to \emph{protections deployed within that software} to secure its assets without relying on external  services.

\softprot comes at a cost. It may add overhead to computation time, used memory and network bandwidth and may negatively impact the user experience.
Mitigating risks from \mate attacks hence means selecting a set of SPs to be deployed on different parts of the application so that the attacker is delayed for a defender-defined time frame without degrading the performance over defender-defined acceptable levels. 

As highlighted in the literature~\cite{basile2023COSE}, SP today often lacks a formal risk analysis and relies heavily on security-through-obscurity. Experts manually select SPs, and their effectiveness and performance are assessed ex-post, i.e., only after deployment. Many challenges remain in achieving automated risk analysis of software.
Formalization and automation are largely required as risk mitigation needs precision, i.e., the repeatability or reproducibility of obtained results~\cite{basile2023COSE}.

Other research highlighted a significant skill gap~\cite{skillgap}: there are not enough experts to protect all software that can benefit from rigorous \softprot; they are costly, hence \softprot is out of reach for SMEs.

Automation is also needed as software vendors face time-to-market pressure.
Every new version of an application needs to be protected. Part of the work on previous versions can probably be reused, but typically, the {\softprot}s at least need to be diversified.
Additionally, software vendors may have to protect many versions, such as ports to different platforms, including mobile devices with limited computational resources.
When proper protection would affect the application's usability due to \softprot{}'s overheads, developers may decide to limit the features on those platforms. For example, media players with DRM will only access low-quality media versions if the platform does not allow full protection.

In this field, substituting human experts is not an easy job. 
The identification of the SP techniques to use, the parts of the software to protect and the configuration of the SPs are left to the `feeling' of the team of experts operating on the code.
Empirical studies aim at modeling the impact of protections against attacks \cite{reganoEmpiric,ceccatoTaxonomy,viticchieEmpirical}. All converge to the need for {formal definitions of potency and resilience}, the criteria introduced by Collberg\etal ~\cite{collberg1997taxonomy}, that allow estimating the effectiveness of \softprot{}s when applied on specific portion of a program.

Moreover, even if human experts were available, latency would still be problematic. Automated tool support can cut the required time and effort.

In this context, our research aims to formalize, automate, and optimize the risk mitigation phase by developing a method to suggest a set of \SPs to apply to different parts of the software to delay attackers without degrading the performance over acceptable levels.

To address these questions and achieve our research goals, the contributions of this work are the following:
\begin{itemize}
    \item a method to compute the effectiveness of protections when applied to software assets' requirements;
    \item a formal model for selecting the optimal \softprot{}s to mitigate risks against the vanilla application, constrained by an overhead threshold;
    \item an approach for finding timely solutions to the above model.
\end{itemize}

The rest of the paper is organized as follows. 
{Section~\ref{sec:overview} presents an overview of a \mate \SP approach to manage \mate attack risks that we previously developed, to frame the novel contributions of this paper.}  
Section~\ref{sec:kb} formally introduces the {constructs} used during the decision process. 
Section~\ref{sec:procedure} describes the model, the algorithms, and the metrics for optimally selecting the mitigations.
Section~\ref{sec:experts} describes how we consulted software protection experts and the inputs they provided for our approach. 
Section~\ref{sec:validation} presents a quantitative and qualitative validation of our models and tools. 
Section~\ref{sec:related} relates our solution to the state of the art. 
Finally, Section~\ref{sec:conclusions} draws conclusions and sketches ideas for future work.

\section{Overview of our approach}
\label{sec:overview}

Protecting software against \mate attacks can be seen as a risk management process. 
The \nist has proposed an \itech systems risk management standard that identifies four main phases \cite{nistSP800-39}:
\begin{enumerate}\itemsep 0pt
\item \emph{risk framing}: to establish the scenario in which the risk must be managed;
\item \emph{risk assessment}: to identify threats against the system assets, vulnerabilities of the system, the harm that may occur if those are exploited, and the likelihood thereof;
\item  \emph{risk mitigation}: to determine and implement appropriate actions to mitigate the risks;
\item \emph{risk monitoring}: to verify that the implemented actions effectively mitigate the risks.
\end{enumerate}

\begin{figure}[t]
	\centering
        \includegraphics{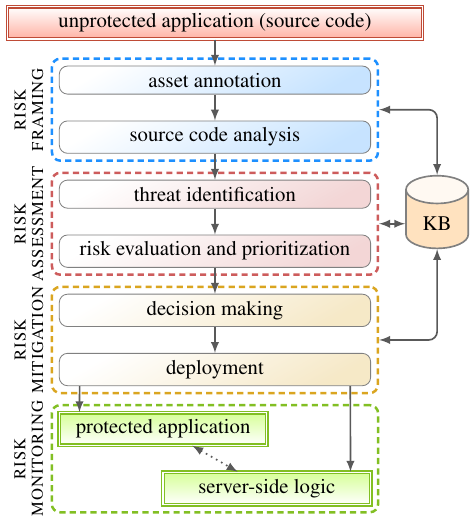}
	\caption{The \esp workflow.}
	\label{fig:workflow}
\end{figure}

Basile\etal ~\cite{basile2023COSE} discussed how this approach can be adopted for \mate \SP. They argued that as much as possible of the four phases should be formalized and automated, and they presented results obtained with a prototype \esp that indeed automates much of the approach. \Cref{fig:workflow} presents the semi-automated workflow of the \esp.
The work presented in this paper is a major contribution of the \esp.\footnote{In the ASPIRE project and some cited papers, the ESP was called the ASPIRE Decision Support System (ADSS).} Its complete code is available,\footnote{\url{https://github.com/daniele-canavese/esp/}} as well as a technical report on its inner workings~\cite{D5.11}, a user manual~\cite{D5.13}, and a demonstration video.\footnote{\url{https://www.youtube.com/watch?v=pl9p5Nqsx_o}}
The \esp is primarily implemented in Java as a set of Eclipse plug-ins with a customized UI. 
It protects software written in C and needs source code access. 
The target users are software developers and \softprot consultants aiming to protect a given application.

\subsection{Risk Framing}
\label{sec:framing}
In the risk framing phase, the \esp user must first annotate the code and data fragments of the C source code of the application that need protection. 
These pragma and attribute annotations identify those fragments as assets and specify their security requirements, which currently include confidentiality and integrity. A formal specification of the annotation language is available online~\cite{D5.11}. 

Using the source code analysis capabilities of the Eclipse C Development Toolkit,\footnote{\url{https://projects.eclipse.org/projects/tools.cdt}} a formal representation of the whole application is then obtained from the source code, according to a software protection meta-model~\cite{reganoMeta}.
In this meta-model, an application is modelled as a hierarchical structure of application parts that can be code regions or data elements (e.g., variables and parameters). 
The relations between those parts in the meta-model are captured in the model instances stored in a \KB, including which code fragments access which data, and the call graph. 

\begin{figure}[tb]
	\centering

	\begin{tikzpicture}[node distance = 7.5mm, scale=0.8, every node/.style = {transform shape}]
	\UMLClass[usecase]{File}{+path}
	\UMLClass[usecase, below = of File]{ApplicationPart}{+name}
	\UMLSimpleClass[usecase, below = of ApplicationPart]{Code}
	\UMLSimpleClass[usecase, left = 15mm of ApplicationPart]{Asset}
	\UMLSimpleClass[usecase, right = 15mm of ApplicationPart]{Datum}
	\UMLEnumeration[generic, below = of Asset]{SecurityRequirement}{+CONFIDENTIALITY\\+INTEGRITY}
	\UMLSimpleEnumeration[generic, above = of Datum]{DatumType}
	\UMLSimpleClass[usecase, below = of Code]{Call}
	\UMLSimpleClass[usecase, below = of Call]{DatumItem}
	
		
	\UMLComposition{ApplicationPart}{File}
	\UMLGeneralization{Asset}{ApplicationPart}
	\UMLGeneralization{$(Code.north) + (2.5mm, 0)$}{$(ApplicationPart.south) + (2.5mm, 0)$}
	\UMLGeneralization{Datum}{ApplicationPart}
	\UMLComposition{$(ApplicationPart.south) + (-2.5mm, 0)$}{$(Code.north) + (-2.5mm, 0)$}
	\UMLAssociation{Asset}{SecurityRequirement}{hasRequirement}
	\UMLAssociation{Datum}{DatumType}{hasType}
	\UMLAssociationRCorner{Code}{$(Datum.south) + (-5mm, 0)$}{accesses}
	\UMLAssociation{$(Code.south) + (-7.5mm, 0)$}{$(Call.north) + (-7.5mm, 0)$}{hasCall}
	\UMLAssociation{$(Call.north) + (7.5mm, 0)$}{$(Code.south) + (7.5mm, 0)$}{hasCallee}
	\UMLAssociation{Call}{DatumItem}{startsWith}
	\UMLAssociationLoop{DatumItem}{isFollowedBy}
	\UMLAssociationRCorner{$(DatumItem.east) + (0, 1mm)$}{$(Datum.south) + (5mm, 0)$}{refersTo}
	\end{tikzpicture}
	\caption{The ApplicationPart class in the \SP meta-model used in the \esp.}
	\label{fig:app-model}
\end{figure}

During the risk framing phase, a catalogue of available {\softprot}s is also collected in the \KB. This includes ordering requirements, restrictions, synergisms, and antagonisms of {\softprot}s. At the time of writing, the \esp supports Tigress, a source code obfuscator developed at the University of Arizona, and the \actc, which automates the deployment of {\softprot} techniques developed in the ASPIRE FP-7 project~\cite{D5.11,D5.13}. Table~\ref{tab:protections} summarizes the {\softprot} techniques supported by the \esp.

\begin{table*}
    \centering
    {\small
    \begin{tabular}{lcccc}
        \toprule
        \multirow{2}{*}{\textsc{protection type}} & \multicolumn{2}{c}{\textsc{requirements}} & \multicolumn{2}{c}{\textsc{tool}} \\ 
        \cmidrule(lr){2-3} \cmidrule(lr){4-5}
        & \textsc{confidentiality} & \textsc{integrity} & \textsc{ACTC} & \textsc{Tigress}\\
        \midrule
            anti-debugging             & \faCheckCircle & \faCheckCircle & \faCheckCircle & \faCircleO\\
            branch functions           & \faCheckCircle & \faCircleO     & \faCheckCircle & \faCircleO\\
            call stack checks          & \faCircleO     & \faCheckCircle & \faCheckCircle & \faCircleO\\
            code mobility              & \faCheckCircle & \faCheckCircle & \faCheckCircle & \faCircleO\\
            code virtualization        & \faCheckCircle & \faCheckCircle &\, {\faCircleO}\tablefootnote{The ACTC provides limited support for code virtualization, meaning that it is not reliably applicable to all code fragments. Hence, the ESP does not consider it a potential protection instance.} & \faCheckCircle\\
            control flow flattening    & \faCheckCircle & \faCircleO     & \faCheckCircle & \faCheckCircle\\
            data obfuscation           & \faCheckCircle & \faCircleO     & \faCheckCircle & \faCheckCircle\\
            opaque predicates          & \faCheckCircle & \faCircleO     & \faCheckCircle & \faCheckCircle\\
            remote attestation         & \faCircleO     & \faCheckCircle & \faCheckCircle & \faCircleO\\
            white-box crypto     & \faCheckCircle & \faCheckCircle & \faCheckCircle & \faCircleO\\
        \bottomrule
    \end{tabular}
    }
    \caption{SPs supported by the \esp, with enforced security requirements and tools used to deploy the SPs. For each tool, we only mark techniques supported on our target platforms, i.e., Android and Linux on ARMv7 processors.}
    \label{tab:protections}
\end{table*}

\subsection{Risk Assessment} 
\label{sec:risk_assessment}
In the risk assessment phase, the threats to the assets are first identified. These threats are represented as a set of \emph{attack paths} that attackers can try to execute.
These paths, in turn, are ordered sequences of atomic attacker tasks called \emph{attack steps}.  
Attack paths are equivalent to attack graphs~\cite{attack_graphs} and can serve to simulate attacks, e.g., with Petri Nets~\cite{petri_nets_attacks}. 
The attack steps that populate our \KB originate from a study and taxonomy by Ceccato\etal~\cite{ceccatoTaxonomy,emse2019} and from data from industrial \softprot experts who participated in the ASPIRE project. In our \esp, the attack steps are rather coarse-grained, such as ``locate the variable using dynamic analysis'' and ``modify the variable statically''. Future work will address this limitation of our \poc implementation.

The attack paths are built via backward chaining as presented in earlier work~\cite{basileOTP,reganoProlog,ReganoPhd}.
An attack step can be executed if its premises are satisfied. It produces the results of its successful execution as conclusions. 
The chaining starts with steps that allow reaching an attacker's final goal (the breach of a security requirement) and stops at steps without any premise. 
The \esp then performs the risk evaluation and risk prioritization by assigning a \emph{risk index} to each identified attack path. 
Every attack step in the \KB is associated with multiple attributes, including the complexity to mount it, the minimum skills required to execute it, the availability of support tools and their usability.  
Additional attributes can be added easily. 
Each attribute assumes a numeric value in a five-valued range.
The values of complexity metrics and software features computed with the available analysis tools on the involved assets are used as modifiers on the attributes to assess the actual risks. 
For instance, an attack step labeled as medium complexity can be downgraded to lower complexity if the asset to compromise has a cyclomatic complexity below some threshold~\cite{mccabe}.
The risk index of an attack path is obtained by aggregating the modified attributes of its steps into a single value. Per attack step, our tool first aggregates all the step's modified attributes into a single attack step risk index. The attack path risk index is then computed from its steps' indices. For more details on this computation, we refer the reader to the existing work from Regano\etal ~\cite{reganoProlog,ReganoPhd}.

\subsection{Risk Mitigation} 
As is commonly done in MATE SP research, we assume that attacks cannot be prevented; they can only be delayed with the help of {\softprot}s. Hence, the mitigation process must select a set of SPs to be applied on parts of the unprotected application such that the attacker will be delayed without degrading the application's performance beyond defender-defined acceptable levels.

\subsubsection{From Risk Index to Software Protection Index}
We model the delaying of attackers as lowering the risk index of their attack paths. The ESP has to find good candidate protection solutions to reduce those risk indices. To identify good candidate solutions, the \esp first searches for \emph{suitable {\softprot}s}, i.e., {\softprot}s that are known qualitatively to impact attributes of the attack steps. 

A solution is an ordered sequence of a number of \SP{}s. In this context, an \SP is not a conceptual construct or method such as ``an opaque predicate''. Instead, it refers to a concrete instantiation, meaning it is a concrete code transformation applied to a specific asset in a specific program by a specific \softprot tool that is configured with specific configuration parameters.

Each such SP is associated with a formula that can alter the attributes of each attack step. 
If a {\softprot} is deployed, the risk index of the attack steps and paths can hence be recomputed to assess the impact of the {\softprot} quantitatively.

For nearly three decades, software metrics have been used to model the strength of software protections quantitatively. Collberg\etal proposed the use of software complexity metrics originating from the domain of software engineering for assessing the potency of protections\cite{collberg1997taxonomy,code_metrics}, and others used quantitative metrics computed on the outputs of software analysis tools to assess the impact of protections on those tools' usefulness for attackers. Examples of the former are Halstead size~\cite{halstead} and cyclomatic complexity~\cite{mccabe}, examples of the latter are points-to-set sizes computed by data flow analysis tools~\cite{foket}, confusion factors of binary code  disassemblers~\cite{linn2003branchFunctions}, and missing edges in function CFGs drawn by GUI disassemblers~\cite{JENS}. The first three example metrics can be considered general-purpose metrics, in the sense that they are relevant to many attack steps and are impacted by many protections. The last two examples are more special-purpose metrics, in the sense that they are relevant for only a limited set of attacker tools and that they are impacted by protections specifically designed for that reason. 

In the ESP, the formulas used to recompute risk indices consider complexity metrics computed on the protected assets' code. 
Additional modifiers are activated when specific combinations of {\softprot}s are applied on the same application part. This way, the 
ESP models the impact of layered and synergetic {\softprot}s when recomputing the risk indices.

Candidate solutions must also meet cost and overhead constraints. 
Our \poc filters candidate {\softprot}s using five overhead criteria: client and server execution time overheads, client and server memory overheads, and network traffic overhead. 

Finally, the \emph{{\softprot} index} associated with a candidate solution is calculated based on the recomputed risk indices of all discovered attack paths against all assets, weighted by the assets' importance. The {\softprot} index is the ESP's instantiation of what is generally called residual risk.

Computing the {\softprot} index by recomputing the risk index requires knowledge of the metrics on the protected application. As applying all candidate solutions would consume an infeasible amount of resources, we have built a \gls{ml} model to estimate the metrics delta after applying specific solutions without having to build the protected application~\cite{reganoMetric}. The ESP's \gls{ml} model has been demonstrated to accurately predict variations of up to three {\softprot}s applied on a single application part. With more {\softprot}s the accuracy starts to decrease significantly. This issue seems to be solvable with larger data sets and more advanced \gls{ml} techniques.

The \esp uses the same predictors to estimate the overheads associated with candidate solutions. Per \softprot and kind of overhead, the \KB stores a formula for estimating the overhead based on complexity metrics computed on the unprotected application.   

\subsubsection{Game-theoretic Optimization Approach}
The possibility, and in practice the necessity, of combining protections greatly increases the solution space. To explore it efficiently and to find (close to) optimal solutions in an acceptable time, the \esp uses a game-theoretic approach, simulating a non-interactive \softprot game. 
In the game, the defender makes one first move, i.e., proposes a candidate solution for protecting all assets. Each proposed solution yields a \emph{base {\softprot} index}, with a positive delta over the risk index of the vanilla application that models the solution's \emph{potency}.

Then, the attacker makes a series of moves corresponding to investments of an imaginary unit of effort in one attack path, which the attacker selects from the paths found in the attack discovery phase. Similarly to how potency-related formulas of the applied {\softprot}s yield a positive delta in the {\softprot} index, we use \emph{resilience}-related formulas that estimate the extent to which invested attack efforts eat away parts of protections and hence of their potency, thus yielding a decreasing {\softprot} index called the \emph{residual {\softprot} index}. This use of resilience aligns with the framing of the potency and resilience terms in a recent survey on SP evaluation methodologies~\cite{desutter2024evaluation}.

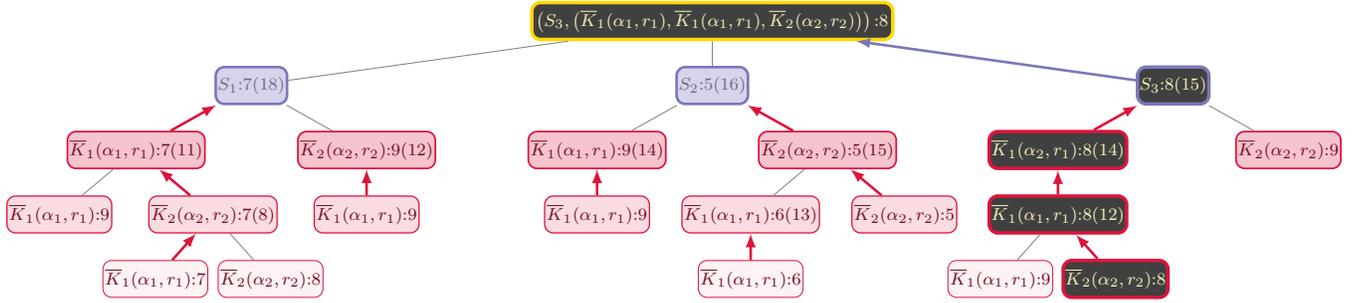
\begin{figure*}[tb]
    \centering
    \resizebox{.99\textwidth}{!}{
    \begin{tikzpicture}[
        level/.style={sibling distance=75mm/#1, level distance=10.5mm},
        every node/.style={rectangle, rounded corners, minimum size=6mm, inner sep=1pt, font=\footnotesize},
        level 1/.style={nodes={very thick, draw=Periwinkle, fill=Periwinkle!25!White, text=Periwinkle!50!Black}},
        level 2/.style={nodes={thick, draw=Crimson, fill=Crimson!25!White, text=Crimson!50!Black}},
        level 3/.style={nodes={semithick, draw=Crimson, fill=Crimson!15!White, text=Crimson!50!Black}},
        level 4/.style={nodes={thin, draw=Crimson, fill=Crimson!5!White, text=Crimson!50!Black}},
        edge from parent/.style={draw, Gray, thin},
        ]
        \node[rectangle, rounded corners, draw=Gold, fill=Black!75, text=Gold!25!White, ultra thick, inner sep=2pt] {$\left( \solution_3, \left( \concreteattackpath_1(\asset_1, \requirement_1), \concreteattackpath_1(\asset_1,\requirement_1), \concreteattackpath_2(\asset_2,\requirement_2) \right) \right){:} 8$}
            child {node {$\solution_1{:} 7 (18)$}
                child {node {$\concreteattackpath_1(\asset_1,\requirement_1){:} 7 (11)$}
                    child {node {$\concreteattackpath_1(\asset_1,\requirement_1){:} 9$}}
                    child {node {$\concreteattackpath_2(\asset_2,\requirement_2){:} 7 (8)$}
                        child {node {$\concreteattackpath_1(\asset_1,\requirement_1){:} 7$} edge from parent[draw=Crimson, very thick, latex-]}
                        child {node {$\concreteattackpath_2(\asset_2,\requirement_2){:} 8$}} edge from parent[draw=Crimson, very thick, latex-]
                    } edge from parent[draw=Crimson, very thick, latex-]
                }
                child {node {$\concreteattackpath_2(\asset_2,\requirement_2){:} 9 (12)$}
                    child {node {$\concreteattackpath_1(\asset_1,\requirement_1){:} 9$} edge from parent[draw=Crimson, very thick, latex-]}
                }
            }
            child {node {$\solution_2{:} 5 (16) $}
                child {node {$\concreteattackpath_1(\asset_1,\requirement_1){:} 9 (14)$}
                    child {node {$\concreteattackpath_1(\asset_1,\requirement_1){:} 9$} edge from parent[draw=Crimson, very thick, latex-]}
                }
                child {node {$\concreteattackpath_2(\asset_2,\requirement_2){:} 5 (15)$}
                    child {node {$\concreteattackpath_1(\asset_1,\requirement_1){:} 6 (13)$}
                        child {node {$\concreteattackpath_1(\asset_1,\requirement_1){:} 6$} edge from parent[draw=Crimson, very thick, latex-]}
                    }
                    child {node {$\concreteattackpath_2(\asset_2,\requirement_2){:} 5$} edge from parent[draw=Crimson, very thick, latex-]} edge from parent[draw=Crimson, very thick, latex-]
                }
            }
            child {node[fill=Black!75, text=Gold!25!White, ultra thick] {$\solution_3{:} 8 (15)$}
                child {node[fill=Black!75, text=Gold!25!White, ultra thick] {$\concreteattackpath_1(\asset_2,\requirement_1){:} 8 (14)$}
                    child {node[fill=Black!75, text=Gold!25!White, ultra thick] {$\concreteattackpath_1(\asset_1,\requirement_1){:} 8 (12)$}
                        child {node {$\concreteattackpath_1(\asset_1,\requirement_1){:} 9$}}
                        child {node[fill=Black!75, text=Gold!25!White, ultra thick] {$\concreteattackpath_2(\asset_2,\requirement_2){:} 8$} edge from parent[draw=Crimson, very thick, latex-]} edge from parent[draw=Crimson, very thick, latex-]
                    } edge from parent[draw=Crimson, very thick, latex-]
                }
                child {node {$\concreteattackpath_2(\asset_2,\requirement_2){:} 9$}} edge from parent[draw=Periwinkle, very thick, latex-]
            }
        ;
    \end{tikzpicture}
    }
    \caption{Search tree example, computed with a mini-max approach and dynamic programming optimizations enabled.}
    \label{fig:tree}
\end{figure*}

Figure~\ref{fig:tree} shows a game tree for a scenario with three candidate solutions $S_1$, $S_2$, and $S_3$; and two possible attack paths $K_1$ and $K_2$ on two assets $\asset_1$ and $\asset_2$ with, in this example, the same security requirements $r$. Each node on the second row models a candidate solution. In a node labeled $s:p(p')$, $s$ is the candidate solution, $p$ is its residual SP index, and $p'$ its base SP index. 

The lower nodes model attack states. For example, the leftmost node on the bottom row models the state reached after a pre-order traversal of the path to that node, i.e., in the state after the attacker has invested in $K_1$ on $\asset_1$, in $K_2$ on $\asset_1$, and in $K_1$ on $\asset_2$. In each node labeled $k(\asset,r):p(p')$, $k$ is the latest attack step, $\asset$ the asset it targets with requirement $r$, $p$ the state's residual SP index considering all succeeding attack steps included in the node's subtrees, and $p'$ the state's residual SP index considering the already executed steps. It can be seen that each additional attack step decreases the $p'$ value as it eats away at the SP index, and that each node's $p$ is the minimum of its childrens' $p'$ values, because the defender makes the worst-case assumption that the attacker will choose the optimal attack path. For leaf nodes, $p=p'$ shows only one residual SP index. 

The directed edges in the graph mark the best attack paths on each candidate solution. The dark nodes mark the best candidate solution, i.e., the one with the highest residual SP index, as well as the best attack path on that solution. The top node of the graph, in essence, summarizes this info. 

In Section~\ref{sec:procedure}, this whole optimization process and the used formulas are discussed in more detail as some of the major contributions of this paper. 

\subsubsection{Deployment of Candidate Solutions}
The ESP then proposes the best solution, possibly with a low number of comparably scoring alternative solutions, to the user, who can make the final selection. The user can then still manually adapt the solution, e.g., fine-tuning some SP configuration parameters, and have the ESP generate the corresponding configuration files for the protection tools. 

At that point, the user can simply invoke those tools (at the moment Tigress and the ACTC, see Section~\ref{sec:framing}) on the application to actually deploy the selected solution in it. The result of this
step (and of the whole workflow) is the protected binary plus source code for the server-side
components for selected online SPs.

Optionally, the ESP can also be asked to deploy additional asset-hiding strategies. In practice, SPs such as obfuscations are never completely stealthy~\cite{reganoL2P}. Instead, they leave fingerprints. If only the assets are obfuscated, those fingerprints facilitate attack steps that aim to locate the assets. The ESP supports three asset-hiding strategies to mitigate this and thus better hide the protected assets. In \emph{fingerprint replication}, {\softprot}s already deployed on assets are also applied to other code parts to replicate the fingerprints such that attackers analyse more parts. With \emph{fingerprint enlargement}, we enlarge the assets' code regions to which the {\softprot}s are deployed to include adjacent regions, so attackers need to process more code per region. With \emph{fingerprint shadowing}, additional {\softprot}s are applied on assets to conceal fingerprints of the chosen {\softprot}s to prevent leaking information on the security requirements. We refer to existing papers~\cite{reganoL2P,basile2023COSE} for more information on this aspect of the ESP's mitigation phase. 

\subsection{Risk Monitoring}
If the selected {\softprot}s include online {\softprot}s such as code mobility~\cite{codeMobility} and reactive remote attestation~\cite{viticchie2016reactive}, the \esp generates all the server-side logic, including the backends that perform the risk monitoring of the released application. This includes the untampered execution as checked with remote attestation and the communication with the code mobility server.

Our \poc does not automatically include the feedback and other monitoring data, such as the number and frequency of detected attacks, compromised applications, and server-side performance issues.
The \KB must be manually updated using GUIs to change risk framing data related to attack exposure and {\softprot} effectiveness. Issues related to insufficient server resources must also be addressed independently; the \esp only provides the logic, not the server configurations.

\section{Formalization - the Knowledge Base}
\label{sec:kb}

The \KB contains the basic structures on which the algorithms operate that Section~\ref{sec:procedure} will describe. These objects, relationships, and properties are based on the meta-model for SP~\cite{reganoMeta} that was mentioned earlier in Section~\ref{sec:framing}. 
This section discusses them in more detail in preparation for Section~\ref{sec:procedure}.

\subsection{Artifacts}
\label{sec:kb:assets}
An \emph{application artifact} $\part $ is a {source code region}, which consists of consecutive source code lines.
The \AAs relevant to the mitigation form the artifact space $\artifacts$. 
Two \AAs are \emph{joint} if they have at least one {source element} in common. This is denoted $\part_1 \joint \part_2 $. Obviously, jointness is commutative ($\part_1 \joint \part_2 \iff \part_2 \joint \part_1$) and idempotent ($\part \joint \part $). In our model, we only consider \AAs that are either completely disjoint, or one is included completely in the other, i.e., they are nested.
We assume that a code normalization pre-pass has been performed, through which variable declarations and other statements, including subexpressions of interest, have all been put on separate lines. Hence variables correspond to proper \AAs.

Variables need special care, however. For each variable, data flow analysis and alias analysis can identify the set of all source code lines that can directly or indirectly (via aliasing pointers) depend on the variable\footnote{We assume that the used SP tools are conservative, in the sense that they will never deploy SPs in ways that alter the application's semantics. The \AAs in our model are only used to determine which transformations will be requested from the tool. The alias analysis used to determine the set of a variable's dependent \AAs can hence be unsound, and in the simplest case even be skipped. While this may yield a suboptimal selection of SPs when the selected ones are not applicable or composable, it cannot yield a non-conservatively protected program. Ideally, the SP tool and the decision support tool of course re-use exactly the same analyses, but this is no requirement.}. 
If this set of a variable's dependent \AAs overlaps with some other \AA, the variable is considered joint with that \AA. 

\subsection{Security Requirements and Assets}

A \emph{security requirement}, denoted with $\requirement$, is taken from the \emph{requirement space} $\requirements$.
Our experiments concerned with two security requirements: \emph{confidentiality} and \emph{integrity}. 
Our approach, however, is extensible and can support any other security requirement.

A \emph{\PO} is a pair $\po{\requirement}{\part}$ that specifies the security requirement $\requirement$ of an \AA $\part $. All the \POs belong to the \emph{\PO space} denoted with $\pos$.

The \emph{assets} of the application are the \AAs that we ultimately need to protect. They appear in at least one \PO pair and form the \emph{asset set} $\assets\subseteq\artifacts$. For ease of notation, we will denote assets as $\asset\in\assets$, to easily tell them from non-assets artifacts $\part\in\artifacts$.
All protection objectives are associated with a non-negative \emph{weight} indicating their importance, which can be retrieved using the function $\weight: \pos \to \nonnegativereals$.

\subsection{Protections}\label{sec:protections}
To protect the security requirements of the assets, the ESP will select \emph{\CSPs} to be deployed on the assets. The \CSPs are the protection instantiations implemented by the used SP tools and configured by the users of the tools. 
Each possible configuration of a supported SP is hence a \CSP $\instance_{i,j}$. 

The total \emph{protection space} is the set $\protections$ of all \CSPs. We partition it into sets $\protection_i=\{\instance_{i,1},...,\instance_{i,n}\}$. Each such set models a single so-called \emph{\ASP}: a set of \CSPs that can be treated as one at certain points in our approach and algorithms. For example, the SP types listed in Table~\ref{tab:protections} correspond to \ASPs. 

This allows for a more concise expression of protection-related information in the \KB, namely per \ASP instead of per \CSP, and it enables optimizations of the algorithms where they can reason per \ASP instead of per CP.\footnote{It are these benefits that determine how to partition $\protections$ into the \ASPs $\protection_i$: if two \CSPs can be considered equivalent with respect to the information and reasoning that is expressed at the  level of \ASPs , they can be added to the same partition. If not, they need to be stored in separate partitions and be considered different \ASPs.}

A \CSP $\instance_{i,j}$ that is deployed on a specific \AA $\part$, is called a \emph{\DSP} and denoted with $\instance_{i,j}(\part)$. All the potential \DSPs from the \emph{\DSP space} $\dpinsts$.

Most SPs can only be deployed on certain types of \AAs. For instance, control flow flattening~\cite{laszloObfuscation} cannot protect variables but only code. We model whether a SP $\protection_i$ is \emph{compatible} with an artifact $\part $ with the Boolean function $\compatible: 2^\protections \times \artifacts \to \{ \true, \false \}$. Furthermore, we model whether a SP affects a security requirement with the function $\enforce: 2^\protections \times \requirements \to \{ \true, \false \}$. 
For instance, control flow flattening helps to preserve confidentiality but not integrity.

Dependencies between \SPs applied to the same \AA are captured with the following relations, which were inspired by the work by Heffner and Collberg~\cite{heffner2004obfuscation}:
\begin{itemize}
    \item \emph{allowed precedence}: $\protection_1 \preceded \protection_2$ indicates that $\protection_1$     \emph{can} precede $\protection_2$, i.e., $\protection_1$ can be applied to some \AA before $\protection_2$;
    \item \emph{required precedence}: $\protection_1 \preceded^{R} \protection_2$ denotes that $\protection_1$ \emph{has to} precede  $\protection_2$;
    \item \emph{forbidden precedence}: $\protection_1 \notpreceded \protection_2$ denotes that $\protection_1$ \emph{cannot} precede  $\protection_2$;
    \item \emph{encouraged precedence}: $\protection_1 \shouldpreceded \protection_2$ indicates that $\protection_1$ \emph{is suggested} to precede $\protection_2$, i.e., this order is particularly beneficial to the \AA's protection. This implies $\protection_1 \preceded \protection_2$;
    \item \emph{discouraged precedence}: $\protection_1 \shouldnotpreceded \protection_2$ denotes that $\protection_1$ \emph{better not} precede $\protection_2$ because this combination negatively impacts the protection. This also implies $\protection_1 \preceded \protection_2$.
\end{itemize}
Note that these relations only restrict the order in which \SP{}s should be applied to some \AA, not whether they need to be applied immediately after each other. For example, applying SPs  $\protection_1$, $\protection_2$ and $\protection_3$ in that order to some \AA is possible if $\protection_1 \shouldpreceded \protection_3$ holds. 
These relations can model various limitations, which may be due to an \SP technique itself or to the used \SP tool. For instance, the fact that some SP that can be applied at most once per asset (e.g., anti-debugging) can be formalized simply as $\protection \notpreceded \protection$.

Dependencies can be expressed as regular expressions~\cite{heffner2004obfuscation} and valid sequences of \CSPs or \ASPs can be generated accordingly. We exploited this property in our implementation. 

\subsection{Solutions}\label{sec:solutions}

A \emph{solution} $\solution$ is an ordered list of \DSPs $\solution = \left( \dpinst{\instance_1}{\part_1}, \dpinst{\instance_2}{\part_2}, \dots \right)$ in the \emph{solution space}  $\solutions$.\footnote{\CSPs have been indexed as $p_{i,j}$ when we were referring to partitions of the protection space. Here, the single indexed $p_{i}$ is used to order \CSPs in a solution.}

The \emph{vanilla solution} is the solution without any \DSPs,  and is represented as $\vanilla \in \solutions$ for any application.

\DSPs and solutions are not inputs of algorithms in Section~\ref{sec:procedure}, they are outputs dynamically computed by our methods.

\subsection{Metrics}\label{sec:metrics}

Our approach and models rely on software metrics for estimating both the effectiveness of protection solutions and their overheads. General-purpose as well as special-purpose metrics are supported, and static metrics such as the mentioned as well as dynamic metrics such as profile information. All the metrics considered by the model are stored in a set $M$.

The optimization process has to examine numerous solutions. Since building binaries and measuring metrics is time-intensive, it is impractical to assess metrics on compiled binaries for all solutions to examine.
Hence, in our model, we introduced an abstract function that predicts the value of the metrics after the application of the solution. 
Formally, the generic function $\metric_m: \solutions \times \artifacts \to \reals$ receives as input a solution and an artifact and returns the metric $m\in M$ of such artifact when it is protected with the solution's \DSPs.

Our current PoC includes the following three general-purpose and three special-purpose metrics:
\begin{itemize}
    \item $\halstead$: the Halstead size of code \AAs, \ie their number of operators and operands~\cite{halstead};
    \item $\cyclomatic$: the cyclomatic complexity of code \AAs, \ie their number of linearly independent paths~\cite{mccabe};
    \item $\instructions$: an AA's number of instructions~\cite{halstead};
    \item $\remoteins$: the number of instructions of an \AA moved to a remote server by SPs such as code mobility~\cite{codeMobility} or client-server code splitting~\cite{barrierslicing,viticchieEmpirical};
    \item $\localins$: the number of instructions of an \AA that has been migrated from the main application process into additional local processes needed for protection purposes, such as the self-debugging code deployed as an anti-debugging protection~\cite{diabloSelfDebugging};
    \item $\guardedins$: the number of instructions of an \AA guarded against tampering by tampering detection techniques such as remote attestation~\cite{viticchie2016reactive}. 
\end{itemize}
Our PoC used the link-time rewriter framework Diablo~\cite{diablo} to measure them on the vanilla application.

Moreover, we have built a pool of ML models that implement the $\metric_m$ function for the PoC supported metrics. Given a \DSP $\dpinst{\instance}{\part}$ and the metrics values for the vanilla $\part $, estimate the metrics values that would be obtained after the \DSP has been applied~\cite{reganoMetric}.

\subsection{Overheads} \label{sec:overheads}
In our model, overheads are real numbers that correspond to ratios of performance metrics before and after protection with a given solution. Multiple performance metrics are supported because multiple types of metrics might be relevant (e.g., space, time, bandwidth) on different application parts (e.g., app initialization vs.\ later phases with real-time requirements, and client-side vs.\ remote server-side in the case of online SPs).  

These overheads can be smaller than one, as some SPs can reduce metrics values. For instance, client-server code splitting can move \AAs to a server~\cite{barrierslicing}, thus reducing the computational resources needed to execute the \AA on the client. 

Formally, the function $\overhead_i: \solutions \times 2^{\artifacts} \to \nonnegativereals$ returns the type $i$ overhead of a deployed solution on a set of artifacts. If this set is the whole program, the total overhead of type $i$ is returned. By specifying these limits for sets of artifacts, the model supports expressing multiple, different constraints on different parts of the application.

The maximum allowed value for the overhead of type $i$ on a set of artifacts $A$ will be denoted by $\maxoverhead_i(A) \in \nonnegativereals$. To specify that we do not care about a specific type of overhead, we can write $\maxoverhead_i(A) = \infty$.

Our current PoC supports five types of overhead, the latter three of which are only considered when online \softprot{}s such as remote attestation are used in a solution: 
 
\begin{enumerate}
    \item client app computation time on sample inputs;
    \item client app memory footprint on those inputs;
    \item server computation time for online SPs; 
    \item server memory footprint for online SPs; 
    \item required network bandwidth.
\end{enumerate}

To estimate the overhead $\overhead_i$ function, our PoC measures the relevant metrics on the vanilla application and estimates them for the candidate solutions, because it cannot generate and measure that much binaries. The formulas used for this estimation were designed to 
estimate an upper bound on the overheads. Alternative techniques for obtaining more precise estimations are certainly useful~\cite{alberto2021towards}, but that is orthogonal to the rest of our approach.

\subsection{Attacks}\label{sec:model:attacks}
An \emph{attack step} will be denoted by $\attackstep$, with all the known attack steps forming the set $\attacksteps$.
When the generic operations implied by an attack step are performed on an \AA $\part\in\mathcal{A}$, we will write $\attackstep(a)$.

An \emph{attack path} $\attackpath(\asset,\requirement)$ that endangers a security requirement $\requirement$ of an asset $\asset$ is an ordered sequence of attack steps that the attacker executes on specific \AAs $\part_i$:  $\attackpath(\asset,\requirement) = \left( \attackstep_1(\part_1), \attackstep_2(\part_2), \dots \right)$. It can be that $\part_i \neq \asset$ when $\part_i $ serves as a pivot to the attacker's goal $\asset $. The \emph{attack path space} $\attackpaths(\asset)$ includes all the attack paths against a specific artifact $\asset$. 

The attack paths $\attackpath$ in $\attackpaths$ and the attack steps $\attackstep$ therein are abstract in the sense that they do not include a notion of effort. In other words, they model virtual attacks in which an attacker has infinite time to execute each attack steps, and therefore, succeeds in every step. 

Since we aim at estimating how SPs delay attacks, we need to incorporate the concept of effort.

Therefore, a \emph{concrete attack path} $\concreteattackpath(\asset,\requirement)$ is the investment of a certain amount of effort in executing a sequence of \emph{concrete attack steps} $\concreteattackstep_i(\asset)$ to endanger the requirement $\requirement$, which are the execution of the attack step $\attackstep_i$ on the \AA $\part_j$ for an imaginary unit of effort.\footnote{It is not useful for optimization purposes to give a precise value of the imaginary unit of effort, as is this just a fixed value to allow comparing the effectiveness of techniques.} The set $\concreteattackpaths(\asset,\requirement)$ denotes the set of all possible concrete attack paths against the requirement $\requirement$ of $\asset$.

The concrete attack paths $\concreteattackpath(\asset,\requirement)$ against an asset are derived from the attack paths $\attackpath(\asset,\requirement)$. 
For an attack path $\attackpath(\asset,\requirement) = \left( \attackstep_1(\part_1), \attackstep_2(\part_2), \dots \right)$, such concrete attack paths are of the form $\concreteattackpath(a,\requirement)=\left( \concreteattackstep_1(\part_1),  \concreteattackstep_1(\part_1), \dots, \concreteattackstep_2(\part_2), \concreteattackstep_2(\part_2), \dots\right)$. When a step $\concreteattackstep_i(\part_j)$ is repeated multiple times, it implies that more than one unit of effort is invested in it. A shorthand for a step $\concreteattackstep_i(\part_j)$ that is repeated $n$ times is to write $ \concreteattackstep_i^n(\part_j)$.

Not all concrete attack paths lead to the violation of a security property. For example, given the abstract attack path $\attackpath(\asset,\requirement) = \left( \attackstep_1(\part_1), \attackstep_2(\part_2)\right)$, investing in the concrete $\left( \concreteattackstep_1(\part_1),  \concreteattackstep_2(\part_2)\right)$ may not be enough for compromising the security properties, while $\left( \concreteattackstep_1(\part_1),\concreteattackstep_1(\part_1),\concreteattackstep_1(\part_1),  \concreteattackstep_2(\part_2),\concreteattackstep_2(\part_2)\right)$ can instead lead to a successful attack.

We model the concept of \emph{probability} of successfully mounting a concrete attack path on a protected asset $\asset $ with the function $\likelihood:\solutions \times  \concreteattackpaths(\asset,\requirement) \to [0, 1]$. 
Such a path's success probability is computed on the success probability of its steps. To that extent, the \emph{concrete attack step probability} $\probability_{\concreteattackstep^n(\part)} \in [0, 1]$ is the probability of successfully executing the concrete attack step $\concreteattackstep$ on the unprotected asset $\asset $ (\ie on the vanilla application) with $n$ imaginary units of effort. The base probability $\probability_{\concreteattackstep(\part)}$ has been explicitly provided by our experts through interviews and questionnaires; the other values of $n$ are obtained with formulas that asymptotically increase from a base probability to $1$. $\pi$ considers the attacker's expertise by changing the base probability. For instance, a guru-level attacker will have a higher probability of success than a script kiddie\footnote{Our approach only requires that the probabilities of successful attack steps are known to compute the protection indices. There is no need to formalize what it actually means to be successful, or even to define successful as the probability of success being 1.}.

In addition, the \emph{mitigation factor} $\mitigation_{\concreteattackstep(\part), \dpinst{\instance}{\part}} \in [0, 1]$ reduces the feasibility of executing the attack step $\concreteattackstep(\part)$ when the \DSP $\dpinst{\instance}{\part}$ is deployed. 

The \emph{synergy factor}  $\synergy_{\concreteattackstep, \dpinst{\instance_i}{\part}, \dpinst{\instance_j}{\part}} \in \nonnegativereals$ is used to model protections' precedences (see Section~\ref{sec:protections}). If $\instance_i \in \protection_i$ and $\instance_j \in \protection_j$, then:

\begin{itemize}
    \item $\synergy_{\concreteattackstep, \dpinst{\instance_i}{\part}, \dpinst{\instance_j}{\part}} > 1$ if $\protection_i \shouldnotpreceded \protection_j$;
    \item $\synergy_{\concreteattackstep, \dpinst{\instance_i}{\part}, \dpinst{\instance_j}{\part}} < 1$ if $\protection_i \shouldpreceded \protection_j$;
    \item $\synergy_{\concreteattackstep, \dpinst{\instance_i}{\part}, \dpinst{\instance_j}{\part}} = 1$ otherwise (\ie only $\protection_i \preceded \protection_j$ holds).
\end{itemize}  
In our PoC, the $\mitigation$ and $\synergy$ are constant values determined using the experts' assessments, even if in the most general case, they may depend on the artifact. 

Finally, given a solution $\solution = \left( \dpinst{\instance_1}{\part_1}, \dots, \dpinst{\instance_n}{\part_n} \right)$, the probability of a concrete attack path $\concreteattackpath(\asset,\requirement)= \left( \concreteattackstep_i^{n_i}(\part_i)\right)_i$
is computed as follows:
\begin{equation*}
    \likelihood \left( \solution, \left( \concreteattackstep_i^{n_i}(\part_i,\requirement)\right)\right)
    =  \prod_{i} \mu(\part_i) \cdot \probability_{\concreteattackstep_i^{n_i}(\part_i)}
\end{equation*}
where $\mu(\part)$ is a function returning the combined effect of the mitigation ($\mitigation$) and synergy factors ($\synergy$) of all the SPs applied to $\part $, so that:
\begin{equation*}
    \mu(\part) =  \prod_{\dpinst{\instance_x}{\part}\in\solution} \mitigation_{\concreteattackstep(\part), \dpinst{\instance_x}{\part}} \cdot \prod_{\dpinst{\instance_y}{\part}\in\solution} \synergy_{\concreteattackstep, \dpinst{\instance_x}{\part}, \dpinst{\instance_y}{\part}}
\end{equation*}

\section{Optimal selection}
\label{sec:procedure}

We leverage a game theoretical procedure to find the optimally protected application. In this scenario, we have two players: the defender, whose goal is to raise the application's security, and the attacker, who tries to diminish it.
Our approach works in two stages: a preparatory stage that precomputes the data structures needed for the second, exploratory stage that searches for the best solutions.

\subsection{Preparatory stage}
\label{sec:preparatory_stage}

The \emph{preparatory stage} is computationally inexpensive but helps improve the optimization process. It executes the algorithms for preparing the list of \DSPs that can be considered for protection, as well as algorithms that partition the \AAs into sets named \CCSs, to speed up the optimization of the \DSPs selection that will be applied on them.

Given the pair $\po{\requirement}{\part}$, which includes a \PO  we want to enforce, we identify the set $\dpinsts_{\po{\requirement}{\part}}$ of all the compatible \DSPs implementing such \PO, which is useful to explore the solution space:
\begin{equation*}
        \dpinsts_{\po{\requirement}{\part}} = \{ \dpinst{\instance}{\part} : \instance \in \protection\!\wedge\!\compatible(\protection, \part)\!\wedge\!\enforce(\protection, \requirement) \}
\end{equation*}

The full \DSP space can be trivially computed as the union of the individual $\dpinsts_{\po{\requirement}{\part}}$.

Attack paths include attack steps that operate on different \AAs (see Section~\ref{sec:model:attacks}). 
{When two attack paths include attack steps operating on the same \AA, one should consider countering all the involved attack steps for deciding how to protect the common \AA.} 
The idea is to partition the \AAs into sets, named the \CCSs, that permit dividing the attack paths so that we can protect artifacts in each set independently. The optimization problem is then split into smaller problems that are more manageable, as the game we propose is, in the worst case, exponential.

Formally, given an asset $\asset$ and all the attack paths 
$\{\attackpath_i(\asset,\requirement)\}_i$ against it that have been determined during the risk assessment phase (see Section~\ref{sec:risk_assessment}), we introduce the function $\art: \artifacts \to 2^\artifacts$, which returns the set of all the artifacts involved in at least one attack step to compromise at least one security property of $\asset$.  In other words, this function returns the \AAs targeted by at least one attack step 
in any of the $\attackpath_i(\asset,\requirement)$.

If $\art(\asset_i) \joint \art(\asset_j)$ does not hold, protecting the two assets will have no interference. 

By contrast, if $\art(\asset_i) \joint \art(\asset_j)$ holds, during the mitigation, the defender may have to deploy protections on common artifacts to mitigate both attack paths.

However, it is not enough to separate the attack paths that have no shared \AAs using $\art$; it is needed to build the closure to be able to partition the optimization problem.
Hence, we use the $\art$ function to partition the artifact space $\artifacts$  into a collection of non-empty \emph{\CCSs} $= \{ \ccs{\assets}{1}, \ccs{\assets}{2}, \dots,\}$ where:
\begin{equation*}
    \ccs{\assets}{i} = \{ \asset_j\in\assets: \exists \asset_k\in\ccs{\assets}{i}, \art(\asset_j) \joint \art(\asset_k) \}.
\end{equation*}

Given this recursive construction, when protecting assets in a \CCS, the assets in all the other \CCSs will not be affected. 
Being partitions, \CCSs satisfy the \emph{coverage} (\ie $\bigcup_i \ccs{\assets}{i} = \assets$) and the \emph{disjointness} (\ie $\bigcap_i \ccs{\assets}{i} = \emptyset$).
Hence, a solution $\solution$ can be split into a set of \emph{partial solutions} $\ccs{\solution}{1}, \ccs{\solution}{2}, \dots$ one for each \CCS.

\subsection{Exploratory stage}\label{sec:exploratory}

The goal of the \emph{exploratory stage} is to find the optimal candidate solutions. 
A \emph{state} $\state = \left( \solution, \attackpathlist_\assets \right)$ is a pair consisting of a solution $\solution$ and an ordered sequence of attack paths $\attackpathlist_\assets = \left( \concreteattackpath(\asset_1,\requirement), \concreteattackpath(\asset_2,\requirement), \dots \right)$ against the assets $\assets$ to protect. 
We will indicate with $\states$ the \emph{state space} so that $\state \in \states$. The simplest state is $(\vanilla, \varnothing)$, which is the vanilla solution without any attacks; note also that the vanilla solution is valid for any application. 

Inspired by game theory, we devised an imaginary turn-based game with two players: the \emph{attacker}, who will invest effort trying a variety of attack paths to compromise the security requirements of the application assets, and the \emph{defender}, who will explore various solutions to protect them.
Unlike traditional games like chess and checkers, however, the first turn is due to the defender, while all the remaining turns are for the attacker. This simulates the situation where a software house tries to publicly release a protected application and attackers have a certain amount of time to try multiple attacks before the value of assets in it decreases to irrelevant values.

\begin{algorithm}[t]
    \KwIn{the attack path space $\attackpaths_\assets$, the solution space $\solutions$, a state $\state = \left( \solution, \attackpathlist_\assets \right)$, and the maximal depth $d$}
    \KwOut{the optimal state $\state'$ of the sub-tree rooted in $\state$ and its protection index $p'$}
    
    \smallskip
    \uIf(\tcp*[f]{the defender's turn}){$\state = \nil$}{\label{ln:defender}
        $p' \gets -\infty$\;
        \ForEach{$\solution \in \solutions$}{
            $\tilde{\state}, \tilde{p} \gets \Explore(\attackpaths_\assets, \solutions, (\solution, \varnothing), d - 1)$\;
            \If{$\tilde{p} > p'$}{
                $p' \gets \tilde{p}$ \;      
                $T' \gets \tilde{T}$\;
            }
        }
    }
    \smallskip
    \uElseIf(\tcp*[f]{a terminal node}){$d = 0$}{\label{ln:leaf}
        $T' \gets \state$\;
        $p' \gets \pindex(T)$\;
    }

    \smallskip
    \uElse(\tcp*[f]{the attacker's turns}) 
    {\label{ln:attacker}
        $p' \gets \infty$\;
        \ForEach{$\attackpath(\asset) \in \attackpaths_\assets$}{
            $\tilde{\state}, \tilde{p} \gets \Explore(\attackpaths_\assets, \solutions, (\solution, \attackpathlist_\assets \cup \attackpath(\asset)), d - 1)$\;
            \If{$\tilde{p} < p'$}{
                $p' \gets \tilde{p}$ \;     
                $T' \gets \tilde{T}$\;
            }
        }
    }
    \smallskip
    \Return{$T'$ \normalfont{and} $p'$}
    \smallskip
    \caption{$\Explore$.}
    \label{al:Explore}
\end{algorithm}

This scenario can be represented with a tree such as the one depicted in Figure~\ref{fig:tree}. 
The first level of the tree contains the (blue) solutions (\ie the defender moves, i.e., the candidate solutions). All the other (red) nodes are concrete attack paths (\ie the attacker moves).
Hence, any path from the root is a state as it includes a specific candidate solution and zero or more concrete attack paths that the attacker may mount to compromise the application when that candidate solution is applied.
Every solution is associated with a base SP index (in parentheses) that is reduced every time the attacker executes a new attack path, yielding a residual SP index with the lowest value being reached at the leaves. 
The (black) \emph{optimal state} contains the \emph{optimal solution} that, after the attack phase, maintains the maximum residual SP index. 
Although the optimal solution is the most interesting information, the other information in the state is the sequence of the most dangerous (black) attack paths, which can also be useful when performing a more educated risk assessment of the application to protect. 

Exploring the graph with a depth-first search algorithm allows one to determine how the solution resists the attack paths and then choose the solutions that resist the best.
The simplest way to explore such trees is to use the recursive Algorithm~\ref{al:Explore} based on the traditional minimax depth-first exploration strategy used in chess programming~\cite{borelMinimax,shannonChess}. 
It receives the attack paths against the assets, the solution space, the state to analyze $\state$, and the maximum remaining depth of the tree to visit (which corresponds to the attack path moves still available to the attacker). It returns the optimal state $\state'$ of the sub-tree rooted in $\state$ and its SP index as outputs.  

The \softprot index is a real number stating how safe a state is. As a rule of thumb, the defender wants to maximize the \softprot index of the application, while the attacker wants to minimize it.

To start the search from the tree root, the first call must be $\Explore(\nil, \attackpaths_\assets, \solutions, d)$, with $d\geq 1$; this will return the optimal state after analyzing the entire tree.
In the initial call, the loop of Lines 3--9 explores all the children of the root note, i.e., all possible moves in the defender's turn. For each of those moves, which correspond to the potential solutions, the recursive call on Line 4 explores the subtree corresponding to the attackers' answers. The answer with the highest residual \softprot index is then selected on lines 5--8 since the defender's goal is to maximize the application's security. 

In the recursive calls, the code on Lines 10--21 is executed to optimize the attackers' answers and the compute the attacker subtrees. The algorithm first checks if the current state is a terminal node at Line~\ref{ln:leaf}. When a terminal node is found, the exploration stops, and the current state and its residual \softprot index are returned.
Otherwise, at Line~\ref{ln:attacker}, the algorithm recursively explores all the attack paths and returns the state with the smallest residual \softprot index, as the attacker's goal is to compromise the application's security.

For example, in the tree in Figure~\ref{fig:tree}, the optimal state is $\left( \solution_3, \left( \concreteattackpath_1(\asset_1,\requirement_1), \concreteattackpath_1(\asset_1,\requirement_1), \concreteattackpath_2(\asset_2,\requirement_2) \right) \right)$ with a protection index of $8$. When the attacker plays, the attack with the lowest residual \softprot index is chosen as the `winning move', and residual \softprot index is propagated upward until it reaches the blue (defender) nodes. 
Dually, the defender's goal is to pick the state with the highest residual \softprot index, and the optimal solution is propagated to the root. 
Algorithm~\ref{al:Explore}'s performance can be vastly improved by adopting a series of well-known dynamic programming optimizations. Namely, we implemented the following techniques:
\begin{itemize}
    \item alpha-beta pruning~\cite{alphabeta}: this variation skips large portions of the tree without impacting the final result;
    \item aspiration windows~\cite{aspirationwindows}: this optimization explores a minimal portion of the tree by guesstimating the protection index range of the optimal state --- this technique is particularly useful when securing a new version of an already analyzed application; thus, when the optimal solution of a similar model is known in advance;
    \item transposition tables~\cite{breuker1997information}: they cache-like objects that store previously computed values related to the protection indices;
    \item futility pruning~\cite{futilitypruning}, extended futility pruning~\cite{extfutilitypruning} and razoring~\cite{razoring}: these reduction techniques aggressively prune forward some states if they seem unpromising.
\end{itemize}
When Algorithm~\ref{al:Explore} is adapted to incorporate these optimizations, the optimized algorithm can decide not to explore some children or sibling nodes, thus making the search tree asymmetric; in these cases, the protection index is computed also in some non-terminal nodes and used to perform an estimation whether or not it is useful to further explore a subtree.

Another simple yet effective optimization exploits the \CCSs. Instead of building a single tree for the whole application, we build a tree for each \CCS $\ccs{\assets}{1}, \ccs{\assets}{2}, \dots$ as each \CCSs are independent pieces of the application. 
The global optimal solution combines together all the optimal solutions for each \CCS, and its residual \softprot index is computed accordingly. 
The $\Explore$ algorithm can be used without any modification; however, one caveat must be reported for our PoC. 
The requirements on overheads are global properties of an entire application and cannot be easily split according to the \CCSs. 
Solutions to this issue are under investigation, in our PoC, we explicitly associated an overhead threshold $\maxoverhead_i$ with each \CCSs. These thresholds were explicitly asked to the industry experts we consulted (as will be discussed in Section~\ref{sec:experts}).

\subsection{Iterating the solution space}
\label{sec:iterating}

The $\Explore$ algorithm requires an efficient manner to iterate through the solution space $\solutions$, which can be too big to fit into memory. 

In our PoC, we decided to explore the solution space $\solutions$ by generating the next solution the mini-max algorithm must process. 
The generation algorithm receives in input a solution $\solution$, the \POs, and \DSPs spaces, and a user-defined integer constant $\maxapis$ specifying the maximum number of \DSPs to use per each \PO. It returns the next candidate solution to explore $\solution'$ or $\nil$ if $\solutions$ has been fully explored.

The algorithm is iterative and does not require storing the entire solution space in memory. However, it needs a starting point. The vanilla solution $\vanilla$ is always a valid starting solution for any application but our approach also took into account the experts' requests to feed the algorithm with their initial candidate solutions.

The function that generates the next solution works in three steps.
\begin{enumerate}
    \item First, it generates a permutation of the \DSPs in the input solution, i.e., it changes the order of the protections in the input solution. 
    It uses the lexicographic permutations with restricted prefixes algorithm~\cite{knuthArt} to correctly take into account protections' precedence and may also exclude discouraged combinations. Further, combinations that exceed the overhead thresholds are discarded.
    
    \item Then, it fuzzes the \DSPs in that permutation, that is, it adds, removes, or replaces some \DSPs with some other ones and checks that all the precedences are satisfied.
    \item Finally, it selects a subset of the \DSPs at the previous step and returns the solution including them (which may contain more, less or the same number of \DSPs of the input solution. 
    
\end{enumerate}
Trivially, the algorithm returns $\nil$, signalling that the solution space has been fully explored\footnote{Even if the solutions are not saved, the used algorithms do not generate duplicates. Hence, we know the generation has been completed by only maintaining counters.}.

\subsection{The Software Protection Index}
\label{sec:procedure:protectionindex}

The \emph{\softprot index} is computed with the function $\pindex: \states \to \reals$. 
Similarly to Collberg's potency, the key property of our protection index is that if a state $\state_1$ is more secure than another state $\state_2$, then $\pindex(\state_1) > \pindex(\state_2)$ must hold, thus permitting us to find the optimal one. 
In addition, the sign of a \softprot index allows us to infer some traits of a state $\state = \left( \solution, \attackpathlist_\assets \right)$:

\begin{itemize}
    \item if $\pindex(\state) = 0$, the state $\state$ is without protections from the attacks in $\attackpathlist_\assets$, as for the vanilla application $\pindex((\vanilla, \varnothing)) = 0$,  
    
    \item if $\pindex(\state) > 0$, the state $\state$ is mitigating the risks against the application, also when the attacker is investing in the attacks in $\attackpathlist_\assets$;

    \item if $\pindex(\state) < 0$, the security of $\state$ is compromised by some attacks in $\attackpathlist_\assets$.
\end{itemize}

Security is a multi-faceted aspect of a protected application; thus, to compute the \softprot index, we decided to use multiple quantifiable security characteristics named \emph{security measures}.

The functions $\measure_i: \assets \times \states \to \mathbb{R}_{\ge 0}$ return the value of the $i$-th security measure of an asset in a particular state. 
We identified four security measures that stem from how \SP{}s work:

\begin{itemize}
    \item $\measure_{CC}$: The \emph{code comprehension} measure estimates how hard it is to understand (local) code. \SPs such as obfuscations increase it, while attacks such as deobfuscation attempt to decrease it. 
    \item $\measure_{CT}$: The \emph{code transfer} measure estimates how much code has been moved to a remote trusted server, thus making it unavailable for reverse engineering on a local machine. For instance, code mobility raises this measure, while attacks on the application's dependence on the remote server (e.g., by reconstructing its funtionality locally) lower it.
    \item $\measure_{TD}$: The \emph{tampering detection} measure evaluates how effective a protection is in detecting an integrity failure. As an example, remote attestation boosts this measure while circumventing or bypassing such a protection reduces it.
    \item $\measure_{TA}$: The \emph{tampering avoidance} measure assesses how effective a protection is in making (static or dynamic) tampering harder. For instance, anti-debugging increases this value, while removing such a technique decreases it.
\end{itemize}

The code comprehension and transfer measures are related to code confidentiality, while tampering detection and avoidance are related to integrity.

All these measures have different relations with the complexity metrics and protections selected, which are captured by our formulas. 
An increase in all the static metrics values, e.g., after obfuscation, has a positive impact on protection, as it is supposed to make code comprehension tasks harder.
Decreasing the code size due to the application of some server-side protections, like client-server code-splitting, has a positive impact on the code transfer measure.
On the other hand, the application of remote attestation is unrelated to the static complexity metrics as it only depends on the technique used and the number and types of attestation checks inserted. 
The general formula for computing the \softprot index of a state $\state = (\solution, \attackpathlist_\assets)$ is:
\begin{equation*}
    \pindex(\state) = \sum_{\asset \in \assets} \left(\weight(\asset) \cdot \left( \sum_i \measure_i(\asset, \state)\right)\right).
\end{equation*}

We computed the $i$-th measure using the equation:
\begin{align*}
    \measure_i(\asset, \state) = \; & \measureweight_i \cdot \measure'_i(\asset, \state) -\\
    & \measuremalus_i \cdot \heaviside(\minmeasure_i - \measure'(\asset, \state)).
\end{align*}

This formula leverages $\measure'_i(\asset, \state)$, an \emph{adjusted measure} that takes into account the effects of \DSPs and attack paths on the asset $\asset$ and the Heaviside step function $\heaviside$.
The adjusted measure is multiplied by $\measureweight_i \in \mathbb{R}_{\ge 0}$, a custom weight introduced to allow us to fine-tune the importance of each measure. 
The second part of the formula subtracts a large constant $\measuremalus_i \in \nonnegativereals$ whenever the adjusted measure is less than $\minmeasure_i \in \nonnegativereals$. 
This subtraction allows marking states for which assets have been breached so that the search algorithm will avoid them\footnote{In chess, this is the equivalent of a checkmate. However, in chess, all checkmate configurations are equivalent, so their score can be set to $-\infty$. By contrast, we need to differentiate a state with a security breach from another state with two breaches, so we cannot set all their \softprot indices to the same value.}.

To compute the adjusted measures, we will make use of the equation
\begin{align*}
    \measure'_i(\asset, \state) = \; & \left( \prod_{\attackpath(a) \in \attackpathlist_\assets} (1 - \likelihood\left( \solution, \attackpath(a) \right) \right) \cdot\\
    & \cdot \measure_i(\asset, (\solution, \varnothing))
\end{align*}
    
This equation uses $\measure_i(\asset, (\solution, \varnothing))$, the $i$-th adjusted security measure computed only on the solution $\solution$ without any attack path. The attack path's influence is instead taken into consideration with the multiplicative factor using the $\likelihood$ function.

To compute $\measure_i(\asset, (\solution, \varnothing))$, we use the utility function $\heaviside'(x) = \heaviside(x) \cdot x$ to simplify some formulas, a variety of complexity metrics and the notion of Collberg's potency~\cite{collberg1997taxonomy} $\potency: \assets \times \solutions \to \reals$. The potency is a value stating how well an artifact is protected and, given the metric $m$ (see Section~\ref{sec:metrics}), it can be expressed as:
\begin{equation*}
    \potency_m(\part, \solution) = \frac{\metric_m(\part, \solution)}{\metric_m(\part, \vanilla)} - 1.
\end{equation*}

Using these definitions, we computed the four adjusted security measures with the following formulas:
\begin{equation*}
    \left\{
    \begin{aligned}
    \measure_{CC}(\asset, (\solution, \varnothing)) = \; & \heaviside'(\potency_{\operatorname{halstead}}(\asset, \solution) +\\
    & \quad\,\, \potency_{\operatorname{cyclomatic}}(\part, \solution))\\
    \measure_{CT}(\asset, (\solution, \varnothing)) = \; & \frac{\metric_{\operatorname{remote.instructions}}(\asset, \solution)}{\metric_{\operatorname{instructions}}(\asset, \varnothing)}\\
    \measure_{TD}(\asset, (\solution, \varnothing)) = \; & \frac{\metric_{\operatorname{guarded.instructions}}(\asset, \solution)}{\metric_{\operatorname{instructions}}(\asset, \solution)}\\
    \measure_{TA}(\asset, (\solution, \varnothing)) = \; & \frac{\metric_{\operatorname{local.instructions}}(\asset, \solution)}{\metric_{\operatorname{instructions}}(\asset, \solution)}.\\
    \end{aligned}
    \right.
\end{equation*}

\section{Expert Consultation}\label{sec:experts}

To instantiate the quantitative optimization approach that we described conceptually in the previous sections, many functions, formulas, factors, parameters, and weights need to be instantiated, i.e., concrete values need to be chosen.
However, known models of MATE attacker behavior~\cite{ceccatoTaxonomy,emse2019} and reverse engineering in general~\cite{remind,Votipka2019} are qualitative. In other words, the literature offers no established, comprehensive quantitative models of how \SPs affect the attacker's performance as needed for our approach. We hence collected the necessary inputs from \SPs developers and industry experts involved in the ASPIRE project; from other experts from the project consortium partners, i.e., that were not performing or assisting the research in the project); and from experts in the Advisory Board.

\subsection{Consulted Experts}

We can broadly distinguish two types of experts that have been consulted through structured interviews.

\subsubsection*{\SPs developers} This category of experts developed SPs, like obfuscation tools, code guards, software attestation techniques, and code mobility, as reported in a project deliverable~\cite{D5.11}.
These experts were primarily asked to answer surveys about the SPs they developed, the security requirements they help preserve, the attackers' activities their techniques impact, and the dependencies with other \SPs, i.e., limitations on composability and potential synergies between them. Moreover, they participated in the surveys related to the definition of the \SP Index function.
    
\subsubsection*{Industry experts} This category of experts includes researchers and practitioners, often with strong backgrounds in offensive tasks and domain knowledge of the use cases on which the project artifacts were evaluated. They worked on designing and analyzing the use case applications to protect, on selecting the proper \SPs as mitigation (as human experts do when no decision support tools are available to automate the selection), on deploying the techniques, and eventually on evaluating empirically whether the protected binaries met the security requirements at an acceptable overhead.
They were asked to answer surveys on the SPs they used for their jobs, following the same approach as the \SPs developers. Moreover, they were {interviewed to acquire empirical information, like expert evaluation of the effectiveness of SPs, complexity of attack steps, and relations among complexity metrics and attack steps and SPs. In short, they were the primary source of information for building the \SP index formulas.

\subsection{Consultation Coverage}

The inputs obtained from the experts cover many areas.

\subsubsection*{Suitability to preserve security requirements (Sec.~\ref{sec:protections})}
\SPs developers' feedback was used to build the $\compatible$ function and, together with Industry Experts' feedback, to define the $\enforce$ function.

\subsubsection*{Relations among \SPs (Sec.~\ref{sec:protections})}
\SPs developers and Industry experts helped to formally model the relations between \SPs to the same \AAs (allowed, required, forbidden, encouraged, discouraged).
These relations have been assessed using \emph{ad hoc} surveys where they were asked to evaluate them on a three-level scale. This information was complemented with the existing literature in the field and with the results of empirical experiments we conducted to assess the efficacy of selected \SPs \cite{reganoEmpiric,viticchie2016reactive}. 
These data helped build the function $\synergy_{\concreteattackstep, \dpinst{\instance_i}{\part}, \dpinst{\instance_j}{\part}}$, i.e., the \emph{synergy factor}.

\subsubsection*{Metrics, SP's and attack complexity relations (Sec.~\ref{sec:metrics})}
SP developers were first asked to indicate the metrics that were 
affected by the application of their SPs. Then, together with Industry experts, they were asked to estimate the impact of variations in the metrics on the complexity of specific attack steps.
These data were used to build the $\measure_i(\asset, \state)$, the parameters $\tau_i$ to relate the importance of the metrics,  and $\measure'_i$ functions.

\subsubsection*{Relations between metrics and SP overheads (Sec.~\ref{sec:overheads})}
\SPs developers answered surveys to determine the association between the metrics and the overheads for the \SPs they owned. This feedback was also used to build the formulas that estimate overheads based on the results of the ML predictors. Moreover, they helped determine the overhead thresholds and their split into \CCSs.

\subsubsection*{Relations between \softprot{}s and attack steps (Sec.~\ref{sec:model:attacks},\ref{sec:metrics})}
    Experts were first asked to evaluate the complexity of individual attack steps, regardless of the presence of SPs. This information served to build the concrete attack steps probability $\probability_{\concreteattackstep^n(\part)}$.
    Then, they were asked to indicate the impact of the SPs in countering the attack steps, which resulted in the mitigation factor $\mitigation_{\concreteattackstep(\part), \dpinst{\instance}{\part}}$.
    We also collected information about attack steps able to weaken specific \softprot{}s, which was used to estimate the effectiveness of SPs and helped build the formulas estimating the probability of successfully mounting an attack path against the security requirement of an asset, i.e., the  $\likelihood$ function. 
    Moreover, this experts' feedback was used to build the resilience-related formulas used during the game-theoretic optimization to estimate the extent to which invested attack efforts eat away parts of the {\softprot} potency, thus decreasing the {\softprot} index.

\section{Validation}
\label{sec:validation}

\begin{table}[tb]
    \centering
    {\small
    \begin{tabular}{lrrrr}
        \toprule
        \multirow{2}{*}{\textsc{use case}} & \multirow{2}{*}{\textsc{\PO}} & \multicolumn{3}{c}{\textsc{CCSs}}\\
        \cmidrule(lr){3-5}
        & & \textsc{count} & \textsc{range} & \textsc{mean}\\
        \midrule
        demo player & 58 & 27 & 1--2 & 1.15\\
        license manager & 59 & 26 & 1--7 & 1.65\\
        OTP generator & 24 & 17 & 1--4 & 1.29\\
        \bottomrule
    \end{tabular}
    }
    \caption{Code correlation sets in our use cases.}
    \label{tab:ccss}
\end{table}

\begin{table}[t]
    \centering
    {\small
    \begin{tabular}{lccc}
        \toprule
        \textsc{stage} & \textsc{algorithm} & \textsc{complexity}\\
        \midrule
        preparatory & \makecell{determine deployed SPs} & linear\\
        \cmidrule(lr){1-3}
        preparatory & \makecell{compute code\\correlation sets} & quadratic\\
        \cmidrule(lr){1-3}
        exploratory & explore search tree & exponential\\
        \bottomrule
    \end{tabular}
    }
    \caption{Computational complexity of the approach.}
    \label{tab:complexity}
\end{table}

The first validation step we have performed is a theoretical complexity analysis. Table~\ref{tab:complexity} summarizes the results; the full results are reported in the Supplemental material, where we also list the pseudo-code for all our relevant algorithms. 
In the worst case, the search tree algorithm has an exponential upper bound complexity. This is the case with and without enabling the dynamic programming optimizations reported in Section~\ref{sec:exploratory}. 

To assess whether this high theoretic complexity impacts the feasibility of the approach, Section~\ref{sec:quantitative} will present an experimental evaluation of the practical usability of the optimization method and the introduced heuristics, showing that in practice, the PoC implementation completes in minutes. Next, Section~\ref{sec:qualitati_analysis} will report a summary of the qualitative evaluation by SP experts.

\subsection{Quantitative Evaluation}\label{sec:quantitative}

We tested our PoC (written in Java) on a virtual machine running on 4~cores of an 11\textsuperscript{th} gen Intel\textregistered{} Core\texttrademark{} i9-11950H@2.60~GHz and 8GB~RAM with Ubuntu~18.04.2 LTS and OpenJDK version~11.0.4~2019-07-16.

Figures~\ref{fig:performance-po}~and~\ref{fig:performance-cap} show the time to find the optimal solution in a variety of applications depending on the \POs and the concrete attack paths. 
The times have been computed on search trees of depths 3, 4, 5, and 6. We have chosen these depth values considering that we used a tree depth of 3 in the qualitative validation reported in Section~\ref{sec:qualitati_analysis}, with which we obtained results considered satisfactory by the SP experts involved in the validation. Furthermore, we considered the numbers of \POs and attack paths ranging from 4 to 512, which we consider reasonable for real-world applications. For example, as reported in Table~\ref{tab:ccss}, the use cases devised to perform the qualitative validation range from 24 to 59 \PO{}s\footnote{Notice that the number of \PO{}s in an application is larger than the number of high-level assets to protect in them. For example, when a high-level asset such as a \emph{license manager} needs to have its integrity protected, the multiple functions that implement it all become individual \PO{}s. The number of high-level assets ranged from 5 to 8 in the use cases used in the qualitative evaluation.}. In addition, we enabled all the supported protections (the complete list is available in the Supplemental material).

\begin{figure}[tb]
    \centering
        \begin{tikzpicture}
            \centering
            \begin{axis}[width=\linewidth, height=0.6\linewidth,
            minor tick num=1,
            scaled x ticks=false, xticklabel style={/pgf/number format/fixed},
            axis lines=left,
            enlarge x limits=0.05,
            enlarge y limits=0.05,
            xticklabel style={rotate=45, anchor=north east},
            xlabel style={yshift=+0.50em},
            ylabel style={yshift=-0.50em},
            legend style={at={(0.05,0.95)}, anchor=north west},
            grid=both,
            grid style={line width=0.5pt, draw=Gray, densely dotted},
            minor tick num=3,
            xtick={4, 32, 64, 128, 256, 512},
            xlabel={\PO count}, ylabel={optimization time [min]}]
                \addlegendimage{empty legend}
                \addplot[color=LimeGreen, mark size=1.5, mark=o, thick] coordinates {
                    (4, 0.0012)
                    (8, 0.001)
                    (16, 0.004666666667)
                    (32, 0.007633333333)
                    (64, 0.02428333333)
                    (128, 0.0952)
                    (256, 0.2212333333)
                    (512, 1.00765)
                };
                \addplot[color=Yellow!50!LimeGreen, mark size=1.5, mark=o, thick] coordinates {
                    (4, 0.0005333333333)
                    (8, 0.0005666666667)
                    (16, 0.01033333333)
                    (32, 0.02703333333)
                    (64, 0.1096)
                    (128, 0.48195)
                    (256, 1.13495)
                    (512, 4.620516667)
                };
                \addplot[color=Orange!50!Yellow, mark size=1.5, mark=o, thick] coordinates {
                    (4, 0.0007)
                    (8, 0.00155)
                    (16, 0.02715)
                    (32, 0.03516666667)
                    (64, 0.3477666667)
                    (128, 1.568016667)
                    (256, 2.955466667)
                    (512, 16.29068333)
                };
                \addplot[color=Crimson!50!Orange, mark size=1.5, mark=o, thick] coordinates {
                    (4, 0.001066666667)
                    (8, 0.0012)
                    (16, 0.1485166667)
                    (32, 0.1391)
                    (64, 0.9725333333)
                    (128, 3.755733333)
                    (256, 8.813033333)
                    (512, 50.13076667)
                };
                \addlegendentry{\hspace{-.6cm}depth};
                \addlegendentry{3};
                \addlegendentry{4};
                \addlegendentry{5};
                \addlegendentry{6};
            \end{axis}
        \end{tikzpicture}
    \caption{Optimization time vs. number of \POs.}
    \label{fig:performance-po}
\end{figure}
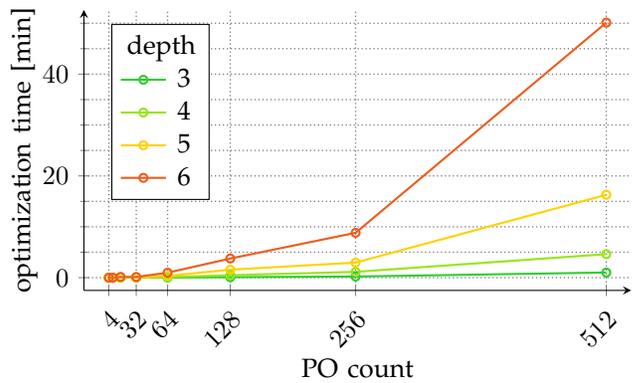

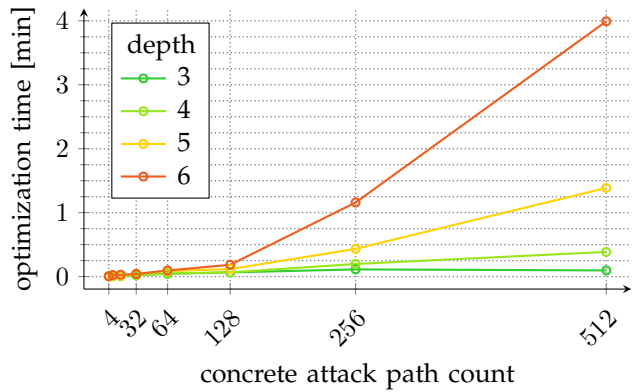
\begin{figure}[tb]
    \centering
        \begin{tikzpicture}
            \centering
            \begin{axis}[width=\linewidth, height=0.6\linewidth,
            minor tick num=1,
            scaled x ticks=false, xticklabel style={/pgf/number format/fixed},
            axis lines=left,
            enlarge x limits=0.05,
            enlarge y limits=0.05,
            xticklabel style={rotate=45, anchor=north east},
            xlabel style={yshift=+0.25em},
            ylabel style={yshift=0em},
            legend style={at={(0.05,0.95)}, anchor=north west},
            grid=both,
            grid style={line width=0.5pt, draw=Gray, densely dotted},
            minor tick num=3,
            xtick={4, 32, 64, 128, 256, 512},
            xlabel={concrete attack path count}, ylabel={optimization time [min]}]
                \addlegendimage{empty legend}
                \addplot[color=LimeGreen, mark size=1.5, mark=o, thick] coordinates {
                    (4, 0.006333333333)
                    (8, 0.007166666667)
                    (16, 0.007633333333)
                    (32, 0.02185)
                    (64, 0.04318333333)
                    (128, 0.06525)
                    (256, 0.1118333333)
                    (512, 0.09606666667)
                };
                \addplot[color=Yellow!50!LimeGreen, mark size=1.5, mark=o, thick] coordinates {
                    (4, 0.004)
                    (8, 0.0053)
                    (16, 0.0111)
                    (32, 0.0198)
                    (64, 0.04901666667)
                    (128, 0.06598333333)
                    (256, 0.1967666667)
                    (512, 0.3856666667)
                };
                \addplot[color=Orange!50!Yellow, mark size=1.5, mark=o, thick] coordinates {
                    (4, 0.004516666667)
                    (8, 0.01366666667)
                    (16, 0.01171666667)
                    (32, 0.04363333333)
                    (64, 0.08646666667)
                    (128, 0.1163333333)
                    (256, 0.4328)
                    (512, 1.3863)
                };
                \addplot[color=Crimson!50!Orange, mark size=1.5, mark=o, thick] coordinates {
                    (4, 0.0038)
                    (8, 0.02356666667)
                    (16, 0.02903333333)
                    (32, 0.03968333333)
                    (64, 0.09498333333)
                    (128, 0.1842833333)
                    (256, 1.158533333)
                    (512, 3.99275)
                };
                \addlegendentry{\hspace{-.6cm}depth};
                \addlegendentry{3};
                \addlegendentry{4};
                \addlegendentry{5};
                \addlegendentry{6};
            \end{axis}
        \end{tikzpicture}
    \caption{Optimization time vs.\ number of attack paths.}
    \label{fig:performance-cap}
\end{figure}

The plots show that the trend is exponential in the number of \POs, concrete attack paths, and search depth. The latter is the most impactful factor on the execution time since we are increasing the tuple length of the concrete attack paths to be analyzed. Interestingly, the \PO count affects the search time more than the concrete attack path count. This is because the number of \POs affects the first tree level and, hence, indirectly, also the whole tree, while the number of concrete attack paths affects all the levels except the first one.

We note that the number of \POs, concrete attack paths, and the search depth primarily influences the computational time. The actual size of the assets (\eg, the source lines of code) does not affect the running time, nor the number of \AAs that are non-assets.

Furthermore, the use of CCSs can mitigate the exponential nature of the search tree algorithm, as this heuristic allows the execution of the algorithm multiple times on smaller sets of \POs. Table~\ref{tab:ccss} shows the size of the CCSs computed on our use cases used for the qualitative validation described in Section~\ref{sec:qualitati_analysis}. The obtained CCS contain at most 7 \POs, with a mean value of less than 2 \POs per CCS in all use cases. Figure~\ref{fig:performance-po} indicates that, for such a small number of \POs, the execution time of the search tree algorithm was in the order of seconds. 

\subsection{Qualitative validation}
\label{sec:qualitati_analysis}
 
This qualitative evaluation reports experts' opinions about our PoC implementation. It was collected from the experts from the industrial partners in the ASPIRE project consortium and its advisory boards. As reported in Section~\ref{sec:overview}, our PoC covers the whole process of protecting a \VA. In this section, we focus on evaluating the mitigation phase, which implements the technique described in Section~\ref{sec:procedure}. 
The complete evaluation of the whole PoC is available in another article~\cite{basile2023COSE} that framed our research in the IT Design Science Framework~\cite{hevner2004design} towards the adoption of the NIST Risk Management Framework (SP800-39~\cite{nistSP800-39}) to progress towards a standardized MATE risk management approach.

The evaluation process objects were three Android apps designed and implemented by the project's industrial partners to represent their commercial software: a One-Time Password generator for home banking apps, an app licensing software, and a \drm-enabled video player for protected content.
These apps included security-sensitive assets in dynamically linked C libraries, which were only made available to academic partners. The high-level descriptions of applications and assets were disclosed in a project deliverable~\cite{D1.06} to confirm they are not toy examples.

Each of the three ASPIRE industrial partners involved two experts to validate their own use cases: one internal expert (i.e., actively involved in the project) and one external expert (i.e., not participating in the project). 
The evaluation was organized into three consecutive phases:
\begin{enumerate}
    \item \emph{Early Internal Expert Assessment}: During the PoC development, the protection owners were involved in evaluating if their individual \SP{}s were used on the proper assets and in the correct way to build solutions. Moreover, internal experts provided continuous feedback on the PoC models, reasoning processes and results. Their feedback drove the PoC development, leading to the alpha version of the PoC, which was tested comprehensively by the internal experts. In particular, they were involved in demos. When the PoC was stable enough, they used it to protect their use case, analyzing and commenting on the results, including the solutions proposed by the PoC and their protection indices. 
    \item \emph{Final Internal Expert Assessment}: Near the end of the project, internal experts were asked to test the PoC's first stable version. They used the PoC's GUI to protect their use case. Moreover, they evaluated the tool's maturity, answering a set of open-ended questions. Such answers were then discussed in multiple calls between the internal experts, the protection owners, the PoC developers and the coordinator (including this paper's authors). The internal experts' comments and suggestions were incorporated into the final version of the PoC.
    \item \emph{Assessment with External Experts}: The PoC's final version was finally tested by external experts, who had never before used the PoC nor had they any information on its internal reasoning processes. They analyzed the results of the PoC execution on their use cases, commenting on the solutions and the individual \SP{}s chosen by the PoC to protect them. They provided their assessment results by answering the questionnaire provided to the internal experts in the previous phase.
\end{enumerate}

{The experts accessed the PoC outputs, an HTML report including the \AAs and assets, the attack paths, and the 10 candidate solutions with the highest solution protection index.}\footnote{These solutions, produced by implementing the technique described in Section 4, are listed in the report as \emph{Level 1 Protections}. The results listed as \emph{Level 2 Protections} are not relevant for this article, since they are produced by an additional reasoning phase of the PoC, where additional protections are applied to non-sensitive code to hide the target application assets and confuse the attacker, following an approach described in a previous publication~\cite{reganoL2P}.} {The PoC reports on the three app use cases, almost identical\footnote{To comply with industrial partners' confidentiality requirements, we renamed code and data identifiers of their use cases.} to the ones analyzed by the experts, are available on GitHub\footnote{\url{https://github.com/daniele-canavese/esp/tree/master/reports}}.} 

The questionnaire answers provided in the second and third phases of the evaluation showed that, in summary, the internal and external experts considered the PoC promising. The degree of automation of the risk management phases, particularly of the mitigation, was perceived as useful to support their daily tasks. 
They noted that the tool could be powerful in the hands of experts due to the high configurability of the internal reasoning processes, which can lead to choices of \SPs for the target application with a quality comparable to a completely manual solution. Conversely, in the hands of software developers without a \SP background and consequently unable to properly fine-tune the PoC parameters, the experts determined that the PoC would not attain the same degree of security for the target application. 

The evaluation of the  candidate solutions proposed by the PoC was positive. The experts highlighted that the \esp's selection of \SP{}s was specifically tailored for the use cases.
Indeed, all the POC's decision processes are based on a formal model of the \AAs constituting the use case code. Thus, attack paths are specific to the target use case. Consequently, since the generation of  candidate solutions considers both the \AAs formal model and the application-specific attack paths, the resulting choice of \SPs is customized for the targeted use case.

The experts also confirmed that the resulting protected applications conserved their original semantics after applying any of the proposed \CSs. 
Furthermore, they agreed on the acceptability of the computational overhead introduced by the chosen \SPs, since the use cases protected with the proposed \CSs were still usable without excessive delays. Also, they reported a high level of obtained asset security since the \SPs included in the \CSs were considered able to protect the use cases appropriately against all the attack paths (see Section~\ref{sec:risk_assessment}) generated by the PoC, and also against the real attacks performed by professional pen testers\footnote{During the ASPIRE project, two external pen testers were tasked to attack the three use cases, each protected with a set of \SP manually chosen by the internal experts. They could not successfully attack the DRM player use case within their available time frame and reported a significant delay in attacking the two other use cases. A report of their activity is available in two ASPIRE public deliverables \cite{D1.06,D4.06}.} and it has been the basis for a journal article \cite{emse2019}.

\section{Related work}
\label{sec:related}

Our work relates to existing work in software protection and risk management.

\subsection{Evaluation of Software Protection Strength}

Our approach's use of complexity metrics is in line with the 1997 proposal of Collberg\etal~\cite{collberg1997taxonomy} to evaluate the potency of protections in terms of complexity metrics. Since then, complexity metrics have been frequently used in literature to evaluate the strength of novel obfuscations~\cite{desutter2024evaluation}.

Our use of complexity metrics to compute protection indices is our implementation of the conceptual 2009 proposal of Nagra and Collberg~\cite{collbergbook} to define potency in terms of extra resources needed for an attacker's analyses to reveal properties of a protected program. Nagra and Collberg define potency in relation to specific analyses to reveal specific properties, which is an improvement over the 1997 definition, but they leave it open how those analyses can be composed of sequences of individual attack steps and how the impact of protections on such compositions should be evaluated. With our approach, we propose a method to specifically solve that issue. 

For each type of attack step, our approach uses distinct formulas in terms of complexity metrics to compute how that specific step's required effort is impacted by the deployed SPs, and how much that attack step can counter that impact, i.e., reduce the protection index. By using distinct formulas for each type of attack step, our approach captures that different metrics are relevant for the different attack steps to be considered. 

By considering only the attack steps that are relevant for the given \PO{}s, i.e., the given assets and their security requirements, and with those distinct formulas, we instantiate the recommendation of De Sutter\etal~\cite{desutter2024evaluation} to evaluate the strength of protections in terms of concrete attacks. By considering both how SPs yield base protection indices, and how attack steps can reduce them to yield residual protection indices, our approach also adopts their recommendation to perform complete evaluations, i.e., to consider both the potency the and resilience of SPs.

\subsection{Automated IT Risk Management}
Research in the automation of risk management procedures in IT systems is rather old, with multiple expert systems for network intrusion detection and auditing being proposed from 1986 onwards~\cite{hoffman1986risk,idesReq}. More recent research mixes expert systems with AI/ML approaches. The work by Depren\etal uses Self Organizing Maps and decision trees for breach detection~\cite{deprenIDS}, feeding these results to an expert system for further interpretation, while the approach by Pan\etal uses neural networks for detecting attacks leveraging zero-day vulnerabilities and an expert system to identify known attacks~\cite{panIDS}. 

A recent survey by Kaur\etal enumerates works for automated risk mitigation in computer networks~\cite{kaur}, distinguishing between approaches for the automated isolation of infected devices and tools for automated recommendation and implementation of risk mitigation procedures~\cite{husak}. 
MATE software protection differs considerably from network security, however. \mate attack modelling needs to include manual tasks and human comprehension of code, which are not considered in network security. For example, in network security, the development of zero-day exploits (using tools also found in the \mate toolbox) is handled as an unpredictable event, which side-steps the complexity of analysing and predicting human activities. This entirely prevents us from reusing of existing assessment models developed for the network security scenario.

\subsection{MATE Software Protection Risk Mitigation}
In a previous paper \cite{basile2023COSE}, we proposed two possible approaches for MATE risk mitigation. The first is performing single-pass mitigation, where a human or a tool is able to find in a single pass the best \SP solution, taking into account also attacks against the protected application. Considering the complexity of the \SP decision process, we deem the automation of this approach unfeasible given the current state of research and the currently available computational resources. The second approach is iterative mitigation, where multiple steps in the \SP decision process are performed. In this approach, a first \SP solution is evaluated on the VA. Then, possible attacks are evaluated on this solution in order to refine it with additional \SP. Multiple rounds of refinement are possible. The game-theoretic procedure presented in Section~\ref{sec:procedure} can be considered a first attempt at automating this procedure since solutions are found iteratively, taking into account the effect of possible attacks against the protected application in terms of a decrease in the protection index. Indeed, this is only an estimation of the actual resilience of selected protections against attacks. In this sense, the approach could be improved by generating multiple versions of the application protected with the \SP solutions with the highest protection index, and automating attack paths found in the risk assessment phase to find the most resilient solution. It should be noted that, given the size of the solution space, it would be practically impossible to perform such a test on all possible solutions. Thus, the game-theoretic approach would still be useful even with an available implementation of such an automated attack procedure.

In industrial practice, companies provide so-called cookbooks with {\softprot} recipes. For each asset, users of their tools are advised to manually select and deploy the prescribed SPs in an iterative, layered fashion as long as the overhead budget allows for additional {\softprot}s. Automated approaches are either overly simplistic or limited to specific types of {\softprot}s, and hence only support specific security requirements. 
Collberg\etal~\cite{collberg1997taxonomy}, and Heffner and Collberg~\cite{heffner2004obfuscation} studied how to decide which obfuscations to deploy in which order and on which fragments given an overhead budget. So did Liu\etal~\cite{obf_optvialangmods,7985664}. They differ in their decision logic and in the metrics they use to measure {\softprot} effectiveness. Importantly, however, their used metrics are fixed and limited to specific program complexity and program obscurity metrics, without adapting them to the identified attack paths.
Coppens\etal  proposed an iterative software diversification approach to counter a concrete form of attack, namely diffing attacks on security patches~\cite{coppens2013feedback}. Their work measured the performance of concrete attack tools to steer diversification and reduce residual risks. All of the mentioned works are limited to obfuscations. In all works, measurements are performed after each round of transformations, much like in the second approach we discussed above. 

To improve the user-friendliness of manually deployed \softprot tools, Brunet\etal proposed composable compiler passes and reporting of deployed transformations~\cite{10.1145/3338503.3357722}. Holder\etal  evaluated which combinations and orderings of obfuscating transformations  yield the most effective overall obfuscation~\cite{obf_evaloptphaseord}. However, they did not discuss the automation of the selection and ordering according to a concrete program and security requirements.

\subsection{Software Protection Tools}
Multiple tools, both commercial and free and open source software (FOSS), are available to automatically \emph{deploy} \SP techniques to protect selected \AA on a target application. 
Our PoC were originally designed in the context of the ASPIRE project, which also developed the ASPIRE Compiler Tool Chain (ACTC), an FOSS toolchain for protecting native ARM Android/Linux libraries~\cite{D1.06}.
Tigress\footnote{\url{https://tigress.wtf}} is another popular automatic \SP tool, that is freely available for research. Tigress is developed by the University of Arizona. This tool performs source-to-source transformations and supports multiple \SP techniques.
The techniques we support for our mitigation phase are the ones of these two tools~\cite{ReganoPhd}. For the protection of natively compiled C/C++ programs, only one additional tool is popular in research according to a recent survey~\cite{desutter2024evaluation}, namely Obfuscator-LLVM~\cite{ieeespro2015-JunodRWM} and more recent derivatives thereof. Those operate on the LLVM Intermediate Representation of the target code to deploy multiple \SP techniques.

Many commercial \SP solutions are available, such as the ones from Irdeto\footnote{\url{https://irdeto.com}}, GuardSquare\footnote{\url{https://www.guardsquare.com}}, VMProtect\footnote{\url{https://vmpsoft.com}} and Oreans (Code Virtualizer\footnote{\url{https://www.oreans.com/CodeVirtualizer.php}} and Themida\footnote{\url{https://www.oreans.com/Themida.php}}). However, scarce information can be derived from commercial descriptions of these tools on their inner workings and the implemented \SP techniques.

There is also research interest~\cite{glamdring,occlum} in automating the deployment of hardware-based \SP{}s, such as Intel SGX and ARM Trustzone. Adapting an application code to support such HW solution is no trivial task, as target application code must comply with multiple requirements (e.g., the use of a modified C Standard Library for SGX-based applications). 

\section{Conclusions and future works}
\label{sec:conclusions}

This paper presented an approach for automatically selecting protections to mitigate risks against assets in software applications. Starting from a vanilla application with annotated assets and previously identified attack paths, the approach employs a game-theoretic method to choose the optimal set of protections, simulating a scenario where a defender uses protection to delay potential attackers. 
The game is solved using a heuristic based on a mini-max depth-first exploration strategy, enhanced with dynamic programming optimizations.
To compare candidate solutions, we introduce the Software Protection Index, which evaluates the effectiveness of protection against specific attack paths. 
We developed a proof-of-concept tool that implements our approach, which experts validated throughout the ASPIRE project. The final assessment confirmed that automated software protection is a viable means for developers and experts to mitigate application risks.

Future work will see technical improvements in the decision-making process. 
Better heuristics in the game-theoretic solver, some inspired by chess, like killer moves, smarter solutions and attack paths visit order, can further improve performance. 

Applying the most recent advances in ML and AI should allow better prediction of metrics used in the computation of the Software Protection Index and overhead estimations. Furthermore, the model for estimating overheads can also be made more precise; we would like to enable the protection experts to express global overheads to be translated into the artifact-specific overheads used by our model.

Moreover, we aim to refine the software protection index to make it a practical yet general implementation of potency and resilience, using more metrics, including the dynamic ones like entropy of memory access patterns and instruction traces, and results from dynamic taint analysis.

Finally, another interesting research area is the automatic generation of more comprehensive attack paths using Large Language Models (LLMs) with Retrieval-Augmented Generation. Indeed, more precise attack paths during the risk assessment phase could help generate even better solutions.

\bibliographystyle{ieeetr}
\bibliography{biblio}

\end{document}